\documentclass[english]{iopart}
\pdfoutput=1
 \usepackage{graphicx}
\usepackage{epsfig}
\input epsf.sty

\newcommand{\beq}{\begin{equation}}
\newcommand{\eeq}{\end{equation}}

\newcommand{\beqr}{\begin{eqnarray}}
\newcommand{\eeqr}{\end{eqnarray}}
\newcommand{\barr}{\begin{array}}
\newcommand{\earr}{\end{array}}
\newcommand{\bal}{\begin{align}}
\newcommand{\eal}{\end{align}} 
\newcommand{\bmu}{\begin{multline}}
\newcommand{\emu}{\end{multline}}

\begin{document}





\noindent
 {\bf \Large Analyticity of the Ising susceptibility: \\ An interpretation}

\author{  M. Assis$^1$, J.L. Jacobsen$^{2,3,4}$, I. Jensen$^1$,  
J-M. Maillard$^5$ and B.M. McCoy$^6$ }
\address{$^1$ School  of Mathematics and Statistics, The
University of Melbourne, VIC 3010, Australia}

\address{${}^2$Laboratoire de physique th\'eorique, D\'epartement de
physique de l'ENS, \'Ecole normale sup\'erieure\\
UPMC Univ.~Paris 06, CNRS, PSL Research University, 75005 Paris,
France}

\address{${}^3$Sorbonne Universit\'es, UPMC Univ.~Paris 06, \'Ecole normale
sup\'erieure, \\
CNRS, Laboratoire de Physique Th{\'e}orique (LPT ENS), 75005
Paris, France}

\address{${}^4$ Institute de Physique Th{\'e}orique,CEA Saclay, 91191
  Gif Sur Yvette, France}
\address{$^5$ LPTMC, UMR 7600 CNRS, 
Universit\'e P. et M. Curie, Paris 6,Tour 23,
 5\`eme \'etage, case 121, 
 4 Place Jussieu, 75252 Paris Cedex 05, France} 
\address{$^6$ CN Yang Institute for Theoretical Physics, Stony Brook
  University, Stony Brook, NY, 11794, USA} 

\begin{abstract}

We discuss the implications of studies of partition function zeros and
equimodular curves for the analytic properties of the Ising model on a
square lattice in a
magnetic field. In particular we consider the dense set of
singularities in the susceptibility of the Ising model at $H=0$ 
found by Nickel and its relation to
the analyticity of the field theory computations of Fonseca and Zamolodchikov.

\end{abstract}
\vspace{.1in}

\section{Introduction}

The magnetic susceptibility at $H=0$ of the two dimensional Ising model on a
square lattice was shown in 1999 by Nickel \cite{nickel1,nickel2} 
to have the remarkable (and unexpected) property that as a function of a complex
temperature variable there is a dense set of singularities\footnote{The emergence of an 
accumulation of singularities had already been seen on resummed series
expansions of {\em anisotropic} Ising models~\cite{ge}.  Here we restrict 
our study to the {\em isotropic} Ising model in a magnetic field.} at the locus of the
zeros of the $H=0$ partition function of the finite size lattice.

On the other hand in 2003 Fonseca and Zamolodchikov \cite{fz} presented a compelling 
scenario, since supported by extensive numerical studies \cite{man1,man2}, 
for the behavior of the Ising model in a magnetic field in the scaling field theory limit 
which assumes analyticity at the locus of singularities.

The compatibility of these two approaches is  an open question which
needs to be understood.

In this paper we investigate this compatibility by means of studying
the dependence on the magnetic field of the temperature zeros 
of the finite size partition 
function and of the equimodular
curves of the corresponding transfer matrix. This will use and extend
the work of \cite{mccoy1}.
It would be highly desirable to treat these questions of  analyticity by
rigorous mathematical methods but, somewhat surprisingly, we will see
that the needed tools do not seem to exist.

 In section 2 we give a precise formulation of the problem. The
 partition function zeros are studied in section 3 and the transfer
 matrix eigenvalues in section 4. In section 5 we use
 these studies to formulate an interpretation which reconciles the
 singularities of Nickel with the analyticity of Fonseca and
 Zamolodchikov. Our conclusions are summarized in section 6.

\section{Formulation}

The isotropic two dimensional Ising model on a square lattice in
the presence of a magnetic field is defined by the
interaction energy
\begin{equation}
{\mathcal E}=-\sum_{j,k}(E\sigma_{j,k}\sigma_{j+1,k}
+E\sigma_{j,k}\sigma_{j,k+1}+H\sigma_{j,k})
\label{interaction}
\end{equation}
where $\sigma_{j,k}=\pm 1$ is the spin at row $j$ and column $k$ and
  the sum is over all spins in a  
lattice of $L_v$ rows and
  $L_h$ columns with either cylindrical or toroidal boundary
  conditions or the boundary conditions of Brascamp-Kunz \cite{bk} 
where on a finite cylinder (with periodic boundary conditions in the
  $L_h$ direction) one end interacts with a fixed row of up spins and the
  other end interacts with an alternating row of up and down spins
  with $L_h$ is even.

The partition function on the $L_v\times L_h$ lattice at
  temperature $T$ is defined as
\begin{equation}
Z_{L_v,L_h}=\sum_{\sigma=\pm 1} e^{-\beta {\mathcal E}}
\end{equation}
where $\beta=1/k_BT$ (with $k_B$ being Boltzmann's constant).
$Z_{L_v,L_h}$ is a polynomial in  the variables 
$u=e^{-2E/k_BT}~~{\rm and}~~x=e^{-2H/k_BT}$.
 However, we note that for appropriate boundary conditions
including Brascamp-Kunz \cite{bk} and toriodal (but not cylindrical) the
dependence is only on $u^2$.
The thermodynamic limit is the limit where $L_v,L_h\rightarrow
\infty$ with $L_v/L_h$ fixed away from zero and infinity. The free energy is
defined in the thermodynamic limit as
\begin{equation}
-F/k_BT=\lim_{L_v,L_h\rightarrow \infty}\frac{1}{L_vL_h}\ln Z_{L_v,L_h}.
\end{equation}
At $H=0$ the free energy of the Ising model is \cite{ons}
\begin{equation}
\hspace{-.2in}-F/k_BT=\frac{1}{2}\ln (2s)+
\frac{1}{8\pi^2}\int_{-\pi}^{\pi}d\theta_1\int_{-\pi}^{\pi}d\theta_2
\ln(s+s^{-1}-\cos\theta_1-\cos \theta_2)
\label{free}
\end{equation}
where 
\begin{equation}
s=\sinh (2E/k_BT)=\frac{u^{-1}-u}{2}.
\end{equation}
This integral has a singularity at a temperature $T_c$ such that $s_c=\pm
1$,  where negative $s$ implies that $E$ is negative and hence that
the system is antiferromagnetic.

For $H=0$ the zeros of the partition function accumulate in the 
thermodynamic limit 
on the circle
\begin{equation}
|s|=1
\end{equation}
which in terms of the variable $u$ becomes the two
circles
\begin{equation}
u=\pm 1+2^{1/2}e^{i\theta}~~~{\rm with }~0\leq \theta< 2\pi
\end{equation}
and the ferromagnetic (antiferromagnetic) critical temperatures are
given by 
\begin{equation}
u_c=\sqrt{2}-1~~{\rm ferromagnetic},~~~u_c=\sqrt{2}+1~~{\rm
  antiferromagnetic}.
\end{equation}
For Brascamp-Kunz boundary conditions all the zeros of the 
partition function for $H=0$ are
exactly on the unit circle at the positions
\begin{equation}
s+s^{-1}=\cos\frac{(2n-1)\pi}{L_h}+\cos\frac{m\pi}{L_v+1}
\label{bkzeros}
\end{equation}
with $1\leq n\leq L_h/2,~1\leq m\leq L_v,$ and $L_h$ even.  

The magnetic susceptibility is given as the second derivative of
the free energy with respect to $H$ as
\begin{equation}
\chi=\frac{\partial M(H)}{\partial H}=k_BT\frac{\partial^2 \ln
  Z}{\partial H^2}.
\end{equation}
In 1999/2000 Nickel \cite{nickel1,nickel2} discovered that in
the thermodynamic limit  for
both $T<T_c$ and $T>T_c$ the susceptibility has an infinite number of
singularities on the circle $|s|=1$ at
\begin{equation}
s_j+s_j^{-1}=\cos (2\pi m/j)+\cos(2\pi n/j)
\label{nickel}
\end{equation}
where
\begin{equation}
0\leq m,n\leq j-1~{\rm with}~m=n=0~{\rm excluded}.
\end{equation}
Here $j$ is a positive integer which is odd for $T>T_c$ and the
singularity at $s_j$ is proportional to
\begin{equation}
\epsilon^{2j(j-1)-1}\ln \epsilon
\end{equation}
where $\epsilon=s-s_j$.
For $T<T_c$ the integer $j$ is even and the singularity at $s_j$ is
proportional to
\begin{equation}
\epsilon^{2j^2-3/2}.
\end{equation}

\section{Partition function zeros}

The partition function depends on the two variables $x$ and $u$ and in
principle should be considered as a polynomial in two
variables. However, here we will consider the dependence on $x$ and $u$ 
separately and not jointly.

\subsection{Dependence on $x$}
The earliest study of partition function zeros is for zeros in the
plane of $x=e^{-2H/k_BT}$ for fixed values of $u=e^{-2E/k_BT}$
where for ferromagnetic interactions $E>0$ and for free, toroidal or
cylindrical boundary conditions Lee and Yang \cite{ly} proved  that the
zeros all lie on the unit circle $|x|=1$
\begin{equation}
Z_{L_v,L_h}(x)=x^{-N/2}\prod_{n=1}^N(x-e^{i\theta_n^{(N)}})
\end{equation}
where $N=L_vL_h$ and $\theta_n^{(N)}$ is real and satisfies
\begin{equation}
\theta^{(N)}_n=-\theta^{(N)}_{N-n}
\end{equation}
and we note that $Z_{L_v,L_h}(x)=Z_{L_v,L_h}(x^{-1})$.
For $T<T_c$, where $0\leq
u<{\sqrt 2}-1,$ the zeros lie on the entire circle $|x|=1$ and for
$T>T_c$, where ${\sqrt 2}-1<u\leq 1,$ the zeros lie on an arc
$x=e^{i\theta}$ where $0<\theta_{LY}\leq \theta \leq
2\pi-\theta_{LY}$.

There have been several numerical studies \cite{kg}-\cite{kim2} of
these zeros and these
studies are all consistent with the limiting statement that, numbering the
zeros as an increasing sequence $\theta^{(N)}_n$ for $1\leq n\leq N$
the limit
\begin{equation}
\lim_{N\rightarrow \infty}N(\theta^{(N)}_{n+1}-\theta^{(N)}_n)
\end{equation}
exists and is non zero. This allows us to define a density for 
${\bar\theta^{(N)}_n}=(\theta_n^{(N)}+\theta^{(N)}_{n+1})/2$ as
\begin{equation}
D({\bar\theta_n})
=\lim_{N\rightarrow \infty}\frac{1}{N(\theta^{(N)}_{n+1}-\theta^{(N)}_n)}
\label{lydensity}
\end{equation}
and for $T>T_c$ this density diverges as $\theta\rightarrow
\theta_{LY}$ and $\theta \rightarrow 2\pi-\theta_{LY}$.

Unfortunately, there are no mathematical proofs for these empirical statements.
For example there is no proof that the density defined by
(\ref{lydensity}) exists and even if it does exist 
the only thing we know about its properties are the values at
$\theta=0$ \cite{yang} and $\pi$ \cite{ly,mccoy3} where
for all $0\leq T<\infty$
\begin{equation}
D(\pi)=\left[
\frac{(1+u^2)^2}{1-u^2}(1+6u^2+u^4)^{-1/2}\right]^{1/4}
\end{equation}
and
\begin{equation}
\hspace{-.8in}D(0)=0~~ {\rm for}~~T>T_c,
~~~D(0)=\left[\frac{1+u^2}{(1-u^2)^2}(1-6u^2+u^4)^{1/2}\right]^{1/4} 
{\rm for}~~T<T_c.
\end{equation}

It is very tempting to write the free energy as
an integral over the density $D(\theta)$ using
\begin{equation}
Z_{L_v,L_h}(x)
=x^{-N/2}\prod_{n=1}^N(x-e^{i\theta^{(N)}_n})
=x^{-N/2}\exp\sum_{n=1}^N\ln(x-e^{i\theta^{(N)}_n})
\end{equation}
so that
\begin{eqnarray}
\hspace{-.5in}
F/k_BT&=&-\lim_{L_vL_h\rightarrow \infty}\frac{1}{L_vL_h}\ln Z_{L_v,L_h}(x)
\nonumber\\
&=&\frac{1}{2}\ln x-\frac{1}{2\pi}\int_{\theta_{LY}}^{2\pi-\theta_{LY}}d\theta
  D(\theta)\ln(x-e^{i\theta})
\end{eqnarray}
 where $(2\pi)^{-1}\int_{\theta_{LY}}^{2\pi-\theta_{LY}}d\theta D(\theta)=1$.
This expression for the free energy is analytic for $|x|\neq 1$. Furthermore
it is universally assumed that on $|x|=1$ the only singularities are at
$x=e^{i\theta_{LY}},e^{i(2\pi -\theta_{LY})}$ for $T>T_c$  and at $x=1$
  for $T<T_c$ \cite{langer} and  the free energy can be analytically 
continued through the arc of zeros on $|x|=1$. This is called the
``standard analyticity assumptions'' in \cite{fz}. However, there is 
absolutely no proof of these assumptions of analyticity.

\subsection{Dependence on $u$ at $H=0~(x=1)$}

\begin{figure}[ht]

\begin{center}
\hspace{0cm} \mbox{
\begin{picture}(350,160)
\put(0,0){\epsfig{file=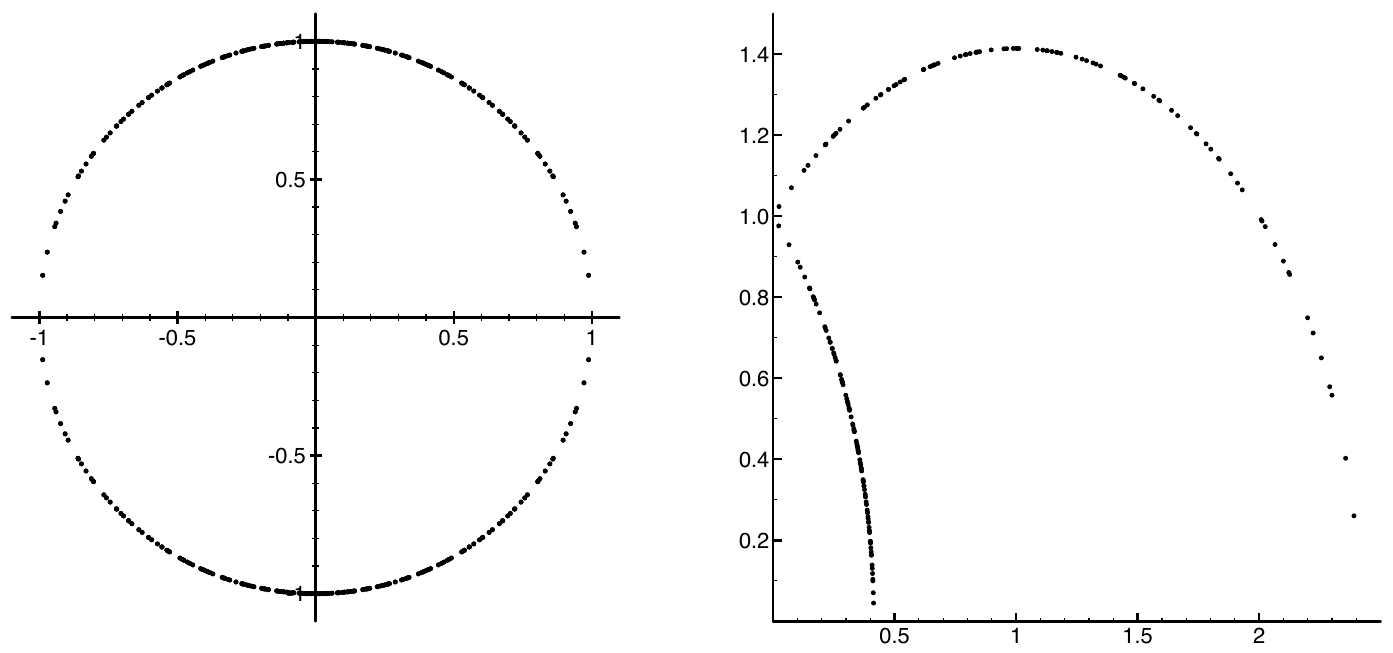,width=12cm,angle=0}}
\end{picture}
}
\end{center}
\caption{Zeros of the isotropic Ising 
model partition function at $H=0~(x=1)$ with
  Brascamp-Kunz boundary conditions for the $20\times 20$ lattice. The
full $s$  plane is plotted on the left. On the right the zeros are
plotted in the $u$ plane; the zeros are on the two circles 
$u=\pm 1+2^{1/2}e^{i\theta}$ and only the first quadrant is shown.
 }
\label{fig1}
\end{figure}

The dependence of the partition function on $u$ for arbitrary fixed $x$
is far more complicated than the dependence on $x$ for fixed $u$. In
particular the zeros in the $u$ plane will not in general lie on
curves  but can fill up areas. The one exceptional case where the zeros
for the finite lattice do lie on curves is when for $H=0$ the lattice has
Brascamp-Kunz boundary conditions. We plot these zeros using
(\ref{bkzeros}) in Figure~\ref{fig1} for the $20\times 20$ lattice in both
the $s$ and the $u$ variable.

Unlike the case of the Lee-Yang zeros in the variable $x$ the zeros in
neither the $s$ nor the $u$ plane have the regular $1/N$ spacing
such that a limiting density defined 
like (\ref{lydensity}) exists. Nevertheless Lu and Wu \cite{luwu}
write the free energy at $H=0$ in the form
\begin{equation}
-F/kT=\frac{1}{2}\ln(4s)+\int_0^{2\pi}d\alpha
 g(\alpha)\ln(s-e^{i\alpha})
\end{equation}
where they ``define'' the density  $g(\alpha)$  
by saying that the
number of zeros in the interval
$[\alpha,\alpha+d\alpha]$
is $L_vL_hg(\alpha)d\alpha$
with
$\int_{0}^{2\pi}d\alpha g(\alpha)=1$.
 
\begin{figure}[h!]

\begin{center}
 \includegraphics[scale=0.28]{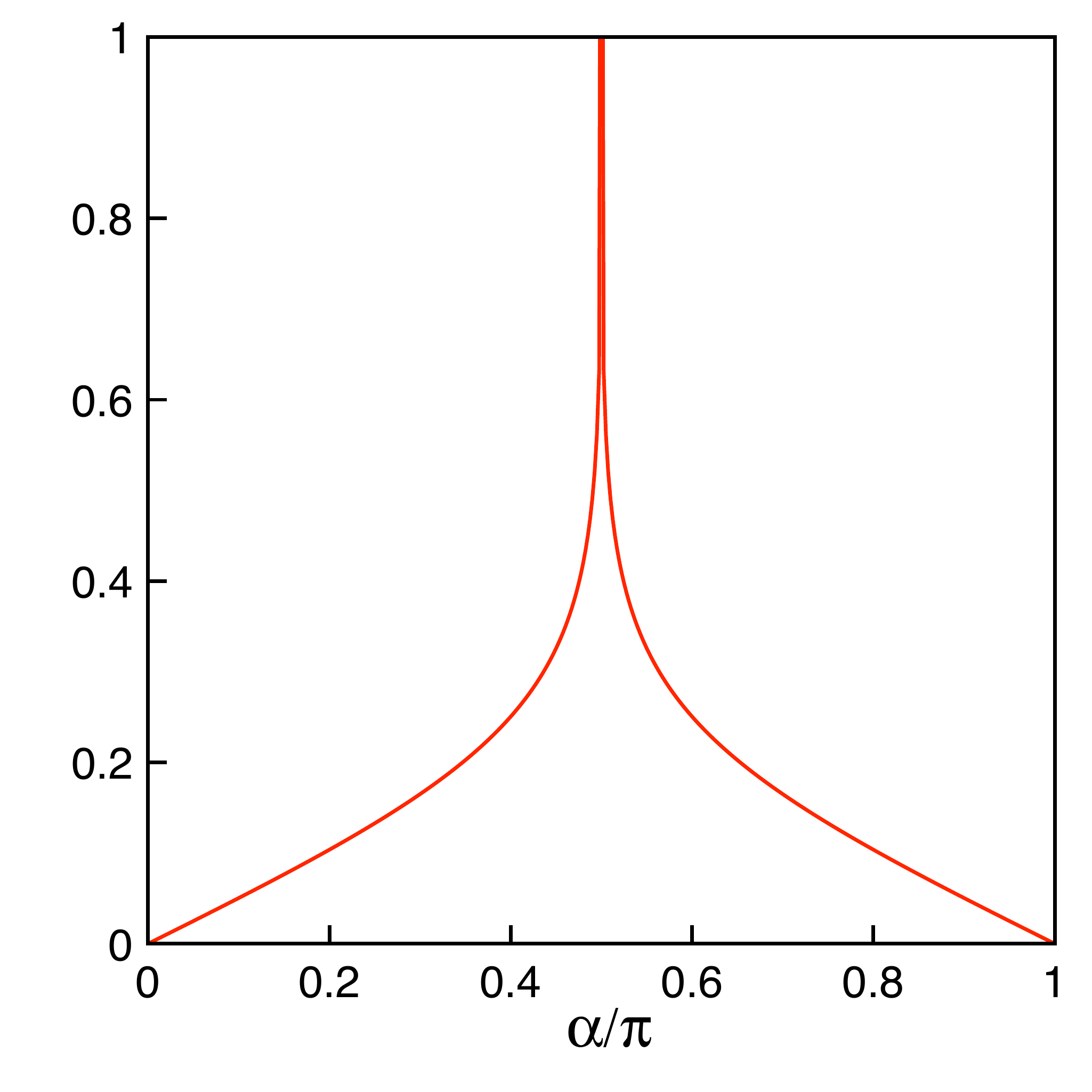}
\end{center}

\vspace{-.1in}
\caption{The density $g(\alpha)$ of Lu and Wu \cite{luwu}.}
\label{fig2}
\end{figure}

This is, of course, a vague statement and is certainly not the same as
(\ref{lydensity}). Then from the two dimensional integral (\ref{free}) 
Lu and Wu  
(and not from the formula for zeros) find
\begin{equation}
g(\alpha)=\frac{|\sin \alpha|}{\pi^2}K(\sin\alpha)
\label{lwdensity}
\end{equation}
where 
\begin{equation}
K(k)=\int_0^{\pi/2}dt (1-k^2\sin^2t)^{-1/2}
\end{equation}
is the complete elliptic integral of the first kind.
We plot this density in Figure~\ref{fig2}.

\begin{figure}[ht]

\begin{center}
\hspace{0cm} \mbox{
\begin{picture}(320,440)
\put(0,0){\epsfig{file=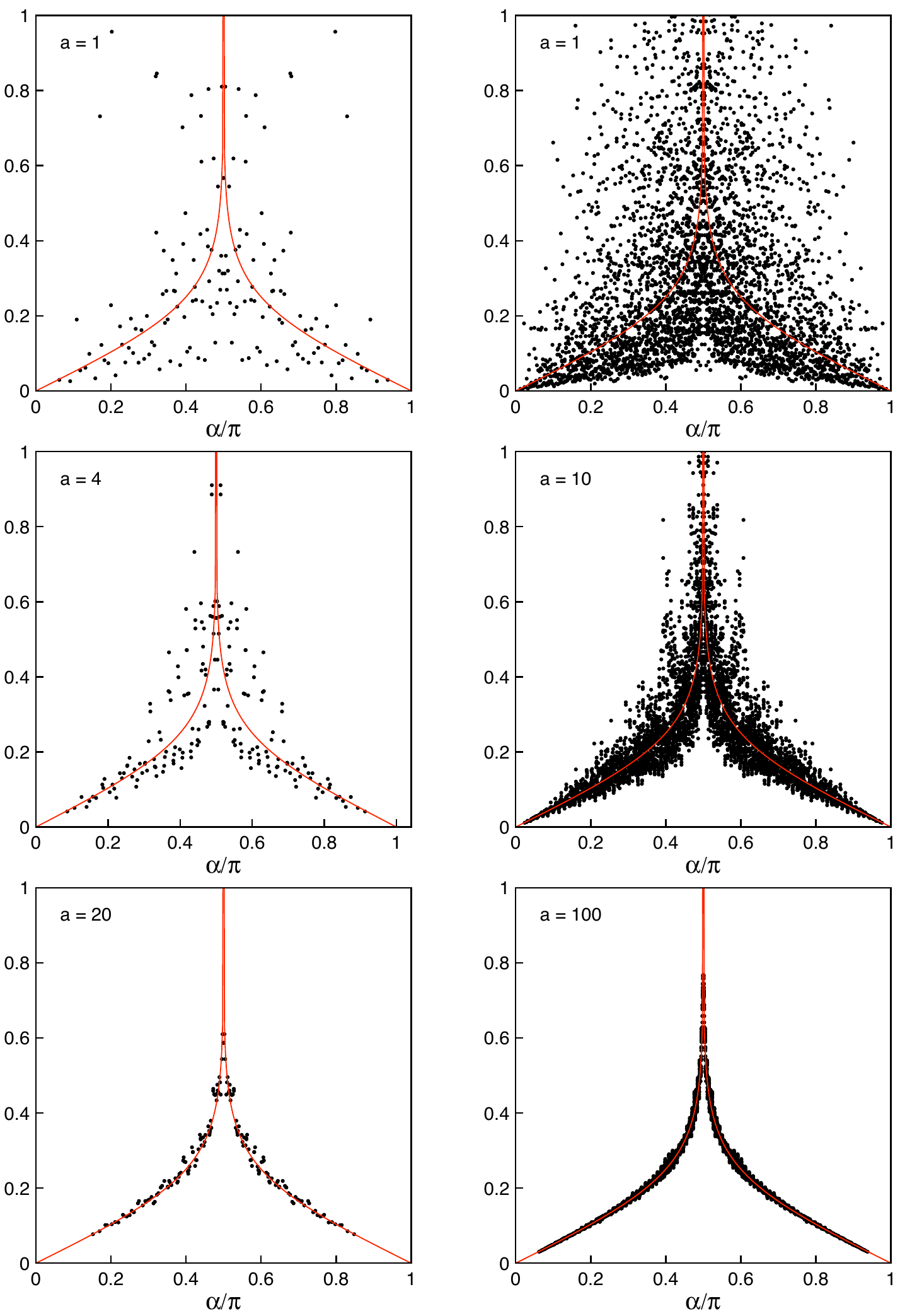,width=11cm,angle=0}}
\end{picture}
}
\end{center}

\caption{Plots of the scale dependent density $g(\alpha;a)_N$ 
for the Brascamp-Kunz
  zeros as a function of the angle $\alpha/\pi$ for the $20\times 20$
  lattice on the left and the $100\times 100$ lattice on the right. In
  the first row $a=1,$ in the second row $a=[L^{1/2}]$ and in the third
  row $a=L=N^{1/2}$. This limiting density (\ref{lwdensity}) of 
\cite{luwu} is shown in red. }
\label{fig3}
\end{figure}

\subsection{Definitions of the density of zeros}

In order to recover the result (\ref{lwdensity}) of \cite{luwu} 
for $g(\alpha)$ from the partition function zeros of (\ref{bkzeros})  we need
to be more precise in the definition of density of zeros. 
There are two slightly different ways to proceed. We can either divide
the circle $s=e^{i\alpha}$ into a set of intervals of equal size and count the
number of zeros in each interval or we can compute the size of an
interval needed to contain exactly a fixed number of zeros. We here
adopt the second method which generalizes (\ref{lydensity})  by defining
\begin{equation}
g(\alpha;a)=\lim_{N\rightarrow \infty}g(\alpha^{(N)}_j;a)_N
\end{equation}
where
\begin{equation}
g(\alpha^{(N)};a)_N=\frac{a}{N(\alpha^{(N)}_{j+a}-\alpha^{(N)}_j)}~~{\rm
  with}
~~a=[cN^p].
\end{equation}
where $[x]$ denotes the integer part of $x$.
If $p=0$ and $c=1$ we recover the density definition
(\ref{lydensity}). If the limit exists for some $p_0<1$ it will
continue to exist for $p>p_0$. The quantity $p_0$ can be called the
scale for which the density exists. 

We examine the existence of these limits for the Brascamp-Kunz zeros 
on the $L\times L $ lattice where $N$ is proportional to $L^2$. In
Figure~\ref{fig3}  we compare for  the $20\times 20$ and $100\times
100$ lattices the scale dependent densities for $a=1,~a=[L^{1/2}]$ and
$a=L=N^{1/2}$. We see for $a=1$ and $a=[L^{1/2}]$ that the limit
  does not appear to exist but the limit does seem to exist 
for $a=L=N^{1/2}.$ Further
  studies reveal that the limit does not exist for $0\leq p<1/2$ 
but does exist for $1/2<p<1$. However, we have no analytic proof
of these numerical observations.

\subsection{Dependence on u for $H>0$}

\begin{figure}[h]

\begin{center}
\hspace{0cm} \mbox{
\begin{picture}(320,440)
\put(0,0){\epsfig{file=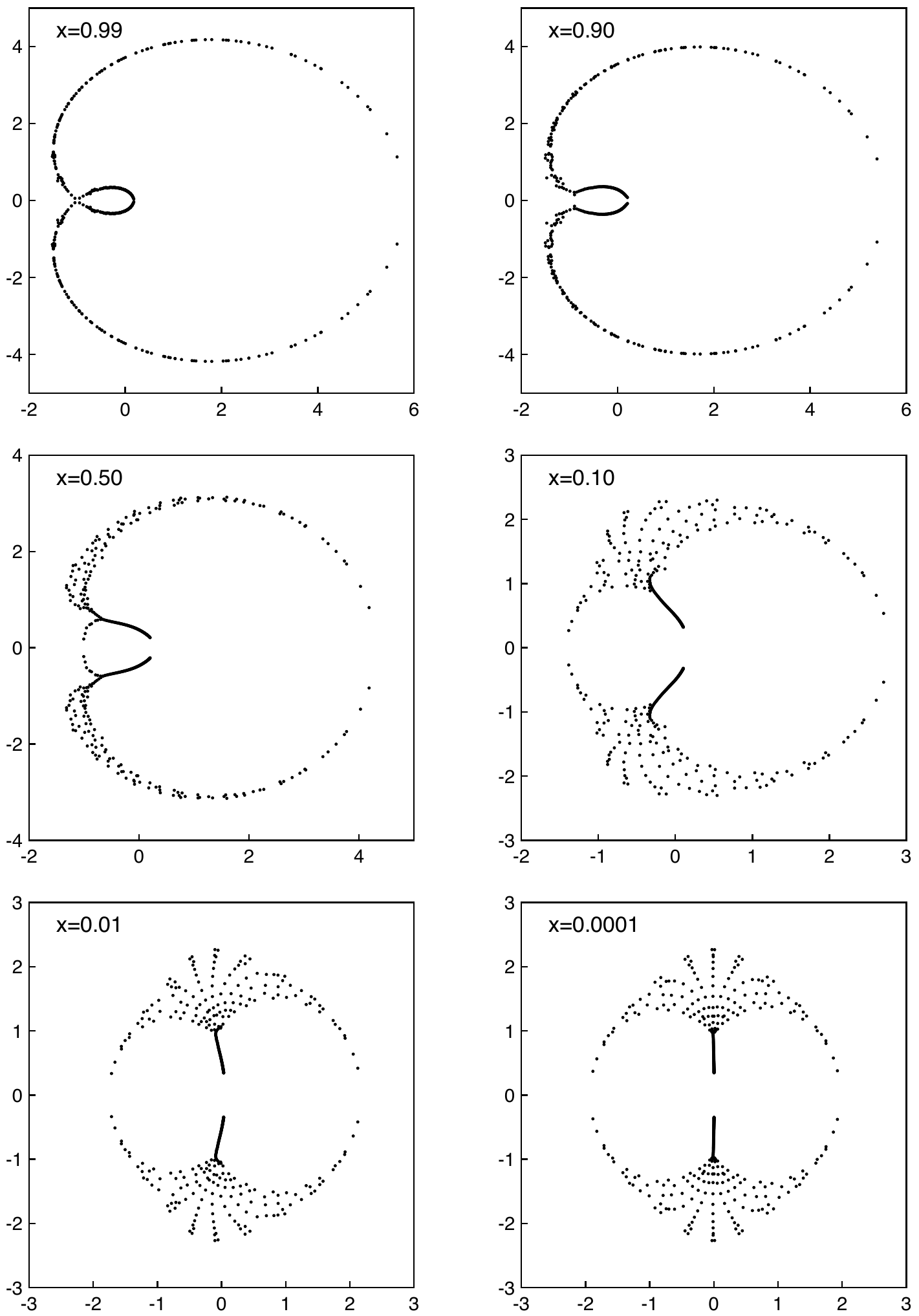,width=11cm,angle=0}}
\end{picture}
}
\end{center}
\caption{Brascamp-Kunz zeros in the plane $y=u^2x^{1/2}$ on the
  $22\times 22$ lattice for values 
of $x=0.99,~0.90,~0.50,~0.10,~0.01,~0.0001$.}
\label{fig4}
\end{figure}

When $H>0$ the free energy is no longer invariant under 
$E\rightarrow -E$  (ie. ferromagnetic $\rightarrow$ antiferromagnetic).
However, for Brascamp-Kunz boundary conditions the partition function
does remain symmetric under $u\rightarrow -u$ and hence 
is a polynomial in $u^2$.
In addition, as the magnetic field $H$ increases the zeros in the $u^2$
plane move to infinity as $x=e^{-2H/k_BT}\rightarrow 0$ so instead of
$u^2$ we consider the rescaled variable 
\begin{equation}
y=u^2x^{1/2}.
\end{equation}
We plot the zeros of the Ising partition function with Brascamp-Kunz
boundary conditions on the $22\times 22$ lattice for several values of
$x$\footnote{The partition function for a given value of $x$
is after multiplication by an appropriate constant a polynomial in $u$ with integer coefficients. The zeros of the partition 
function can then be calculated numerically (to any desired accuracy) using  
root finders such as {\tt MPSolve} \cite{Bini00} or {\tt Eigensolve} \cite{Fortune02}.}  in Figure~\ref{fig4}.
These extend the earlier work of Matveev and
Shrock \cite{shrock1} on $7\times 8$ lattices with helical boundary
conditions and Kim \cite{kim} on $14\times 14$ lattices with
cylindrical boundary conditions.

\begin{figure}[h!]

\begin{center}
\hspace{0cm} \mbox{
\begin{picture}(360,180)
\put(0,0){\includegraphics[scale=0.3]{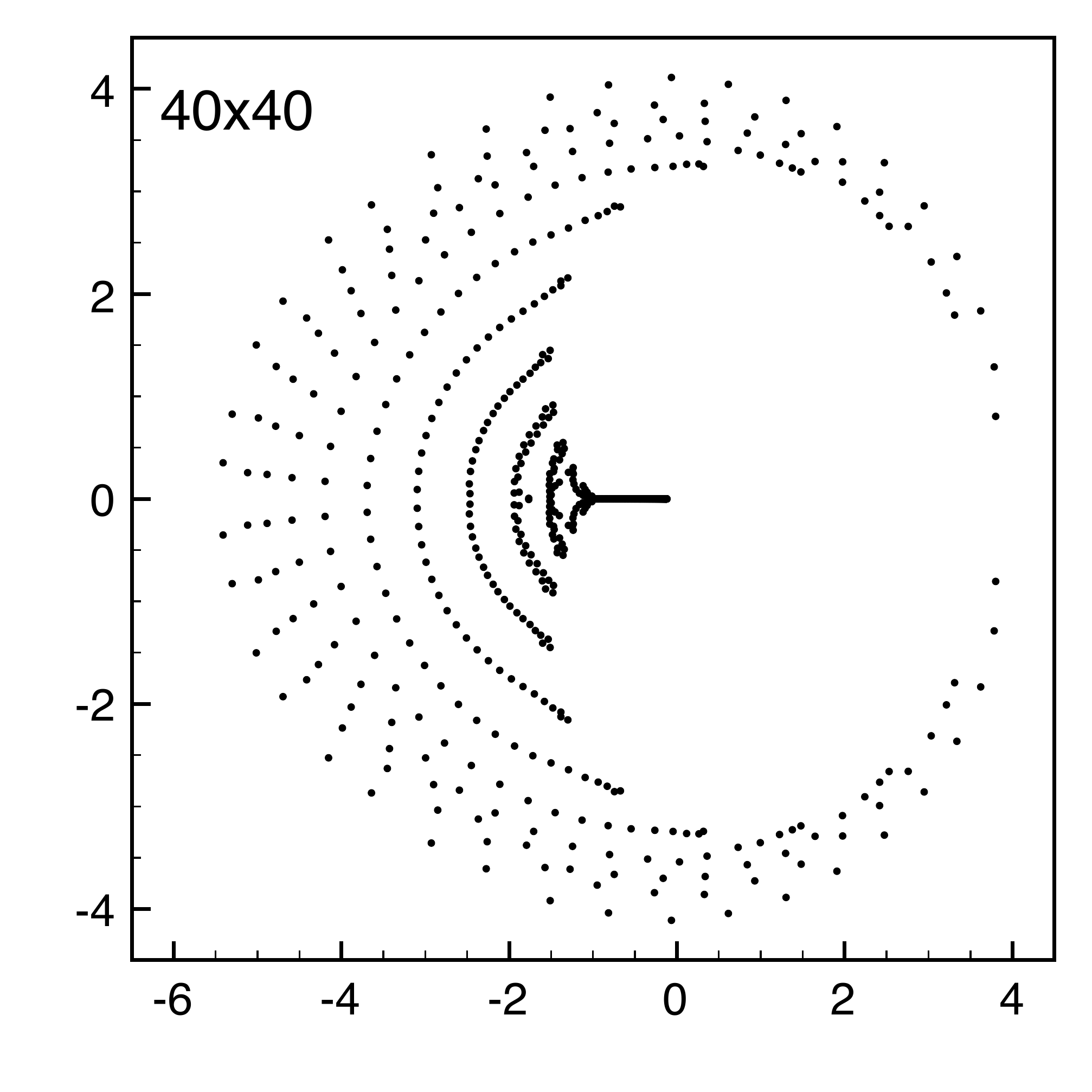}}
\put(180,0){\includegraphics[scale=0.30]{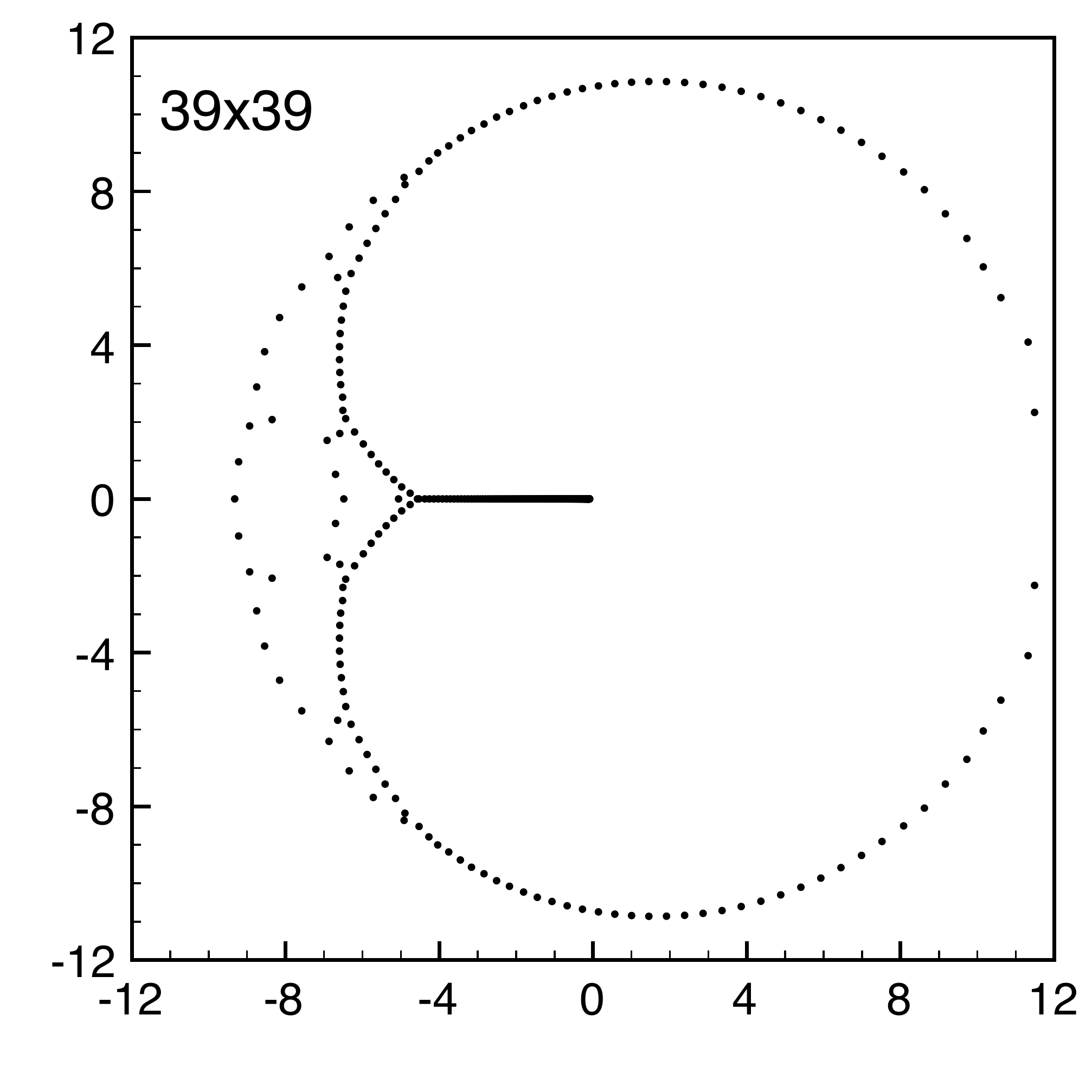}}
\end{picture}
}
\end{center}

\caption{Comparison in the complex fugacity plane $z$ of the 
zeros of the partition function  with 
cylindrical boundary of hard squares on the $40\times 40$ lattice 
to hard hexagons on the $39\times 39$ lattice taken from Figure 2 of
ref. \cite{mccoy1}.}
\label{fig5}
\end{figure}

It is quite clear from these plots that as $H\rightarrow
\infty~(x\rightarrow 0)$ the zeros become symmetric under
$y\rightarrow -y$. This limiting case of the Ising model on
the isotropic square lattice is the hard
square system at fugacity
\begin{equation}
z=y^2
\end{equation}
which has been studied in \cite{mccoy1} for cylindrical boundary
conditions on the $40\times 40 $ lattice. We
plot these zeros in Figure~\ref{fig5} along with the similar plot
for hard hexagons on the $39\times 39$ lattice for comparison.

\clearpage

It is strikingly obvious that as $H$ increases from zero that the
inner and outer loops in Figure~\ref{fig4} behave in drastically
different ways. The  inner loop in Figure~\ref{fig4} 
which separates the disordered from the ferromagnetic 
ordered phase smoothly becomes the line $-1\leq z \leq z_d$ of hard squares 
whereas the outer loop does not remain a curve and spreads out into a
two dimensional area. These two regions must be treated separately.

\subsection{The inner loop zeros}

\begin{figure}[ht]

\begin{center}
\hspace{0cm} \mbox{
\begin{picture}(300,450)
\put(0,0){\epsfig{file=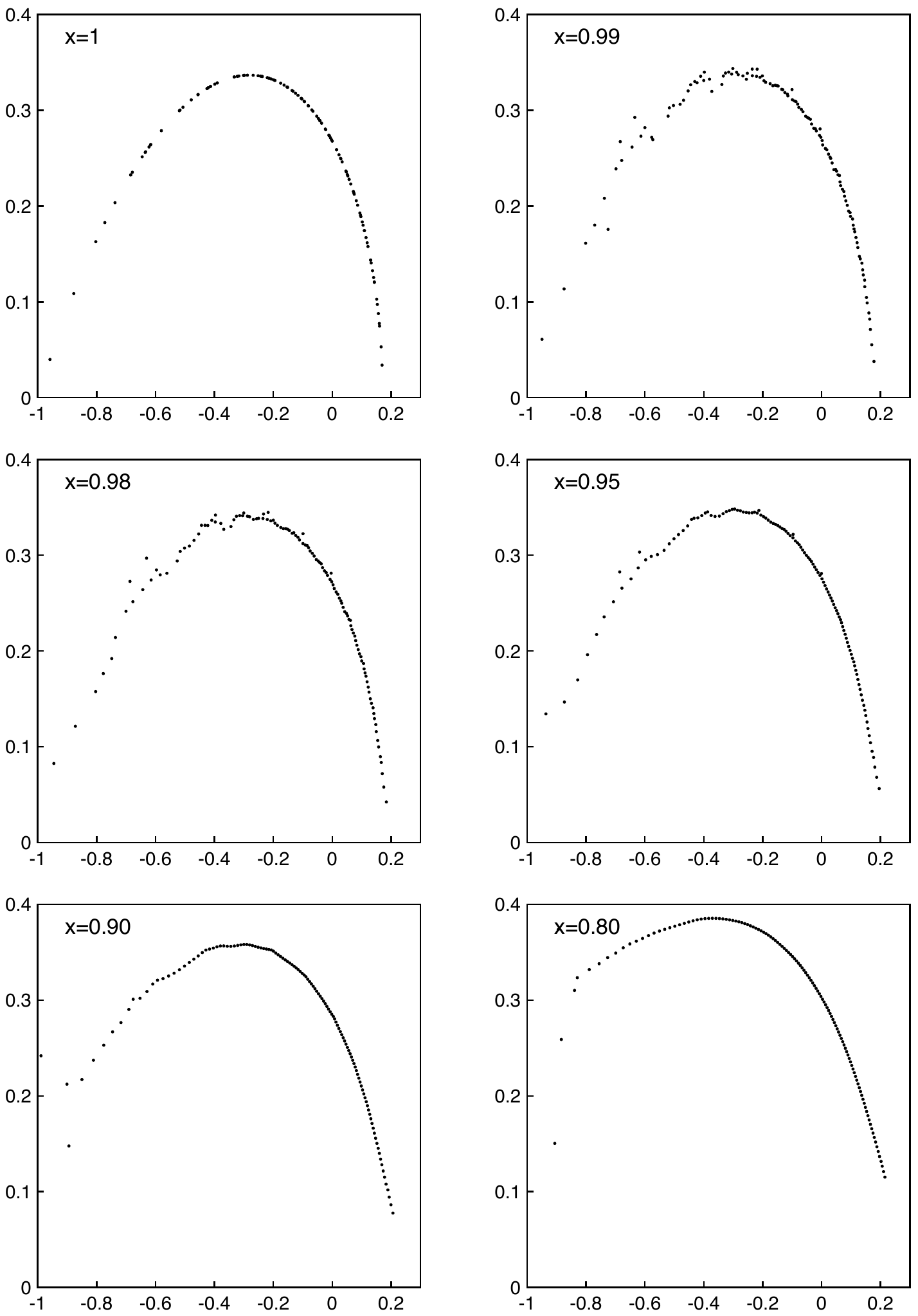,width=11cm,angle=0}}
\end{picture}
}
\end{center}
\caption{Partition function zeros for the $22\times 22$ lattice with
  Brascamp-Kunz boundary conditions on the inner loop in the plane 
$y=u^2x^{1/2}$ for $x=1.0,~0.99,~0.98,~0.95.~0.90,~0.80$}
\label{fig6}
\end{figure}

To study the inner loop zeros in more detail we plot them on an 
expanded scale in Figure~\ref{fig6} for a $22\times 22$ lattice.

\begin{figure}[h]

\begin{center}
\hspace{0cm} \mbox{
\begin{picture}(300,450)
\put(0,0){\epsfig{file=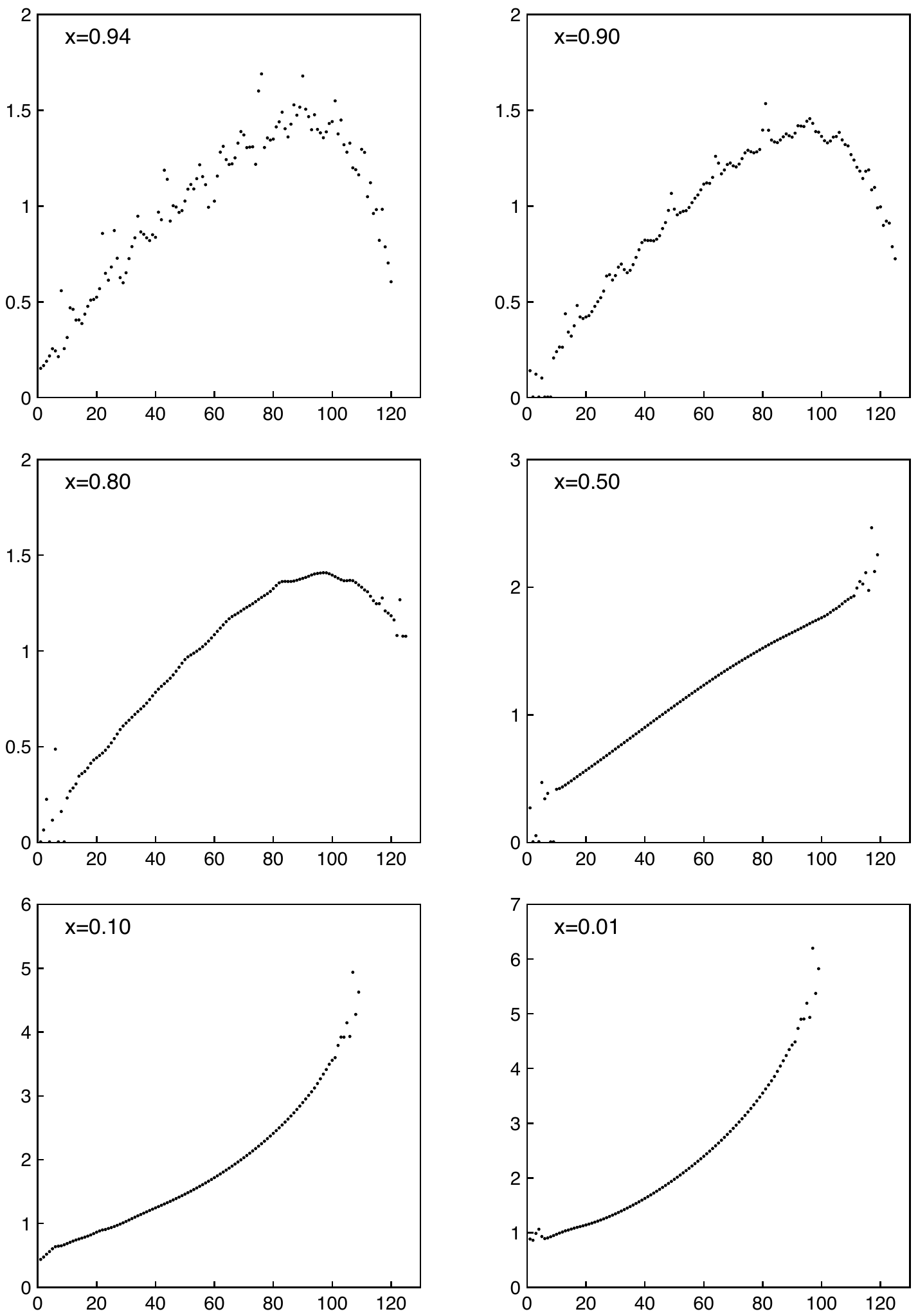,width=11cm,angle=0}}
\end{picture}
}
\end{center}
\caption{The nearest neighbor density of zeros (\ref{bkdensity}) of
  the $22\times 22$ lattice with Brascamp-Kunz boundary conditions
in the plane $y=u^2x^{1/2}$  for
$x=0.94,~0.90,~0.80,~0.50,~0.10,~0.01$ versus the
the index $j$ .}
\label{fig7}
\end{figure}

These plots make it abundantly clear that there is a sharp change in
behavior which sets in as soon as $H$ is increased from zero and that
this transition has been completed for $x<0.95$. In the region
$0.95\leq x <1$ the deviations from a smooth curve become sufficiently
large that a one dimensional density formula becomes
inappropriate. Furthermore it is likely that the structure in this
region will change with increasing lattice size. However, for $x<0.95$
the locus of zeros has become quite smooth and we can consider a
density function
\begin{equation}
D(y_j)=\frac{1}{N|y_{j+1}-y_j|}
\label{bkdensity}
\end{equation}
where $y_j$ is the position of the $j^{th}$ zero as measured from the
endpoint on the right and $N$ is the number of zeros on the inner loop.
We plot this density in Figure~\ref{fig7} versus
the index $j$.

 For $x>0.90$ it is clear from Figure~\ref{fig7} that the nearest neighbor
 density is not smooth for $L=22$. This connects with the behavior 
already seen for $H=0$. However, for $x\leq 0.8$ the nearest neighbor 
density is very smooth except at the rightmost end and  the
 spacing of zeros behaves for large $N$ as $1/N$ which is what was
 observed for hard squares and hexagons in \cite{mccoy1}.

Universality suggests that for sufficiently large $N$ the density at
the right-hand endpoint should diverge for all $x<1$. This is more or less seen
qualitatively in Figure~\ref{fig7} for $x<0.5$ and in the hard square limit
an exponent of $1/6$ was estimated in \cite{mccoy1} from the data of
the $40\times 40$ lattice. However, it is not possible to extract an
accurate exponent of divergence from the data shown in Figure~\ref{fig7}.

\subsection{Outer loop zeros}

The zeros on the outer loop behave very differently from the inner
loop zeros. Instead of the zeros of $H=0$ changing their spacing to
the density function (\ref{bkdensity}) the zeros have spread out into
an area which grows as $H$ increases. It may be conjectured that this
spreading into an area happens for the entire outer loop but for any
finite size lattice there will always be a region near the real axis
where this effect cannot be resolved.

\subsection{Toroidal and cylindrical boundary conditions}

\begin{figure}[ht]

\begin{center}

\hspace{0cm} \mbox{
\begin{picture}(260,400)
\put(0,0){\epsfig{file=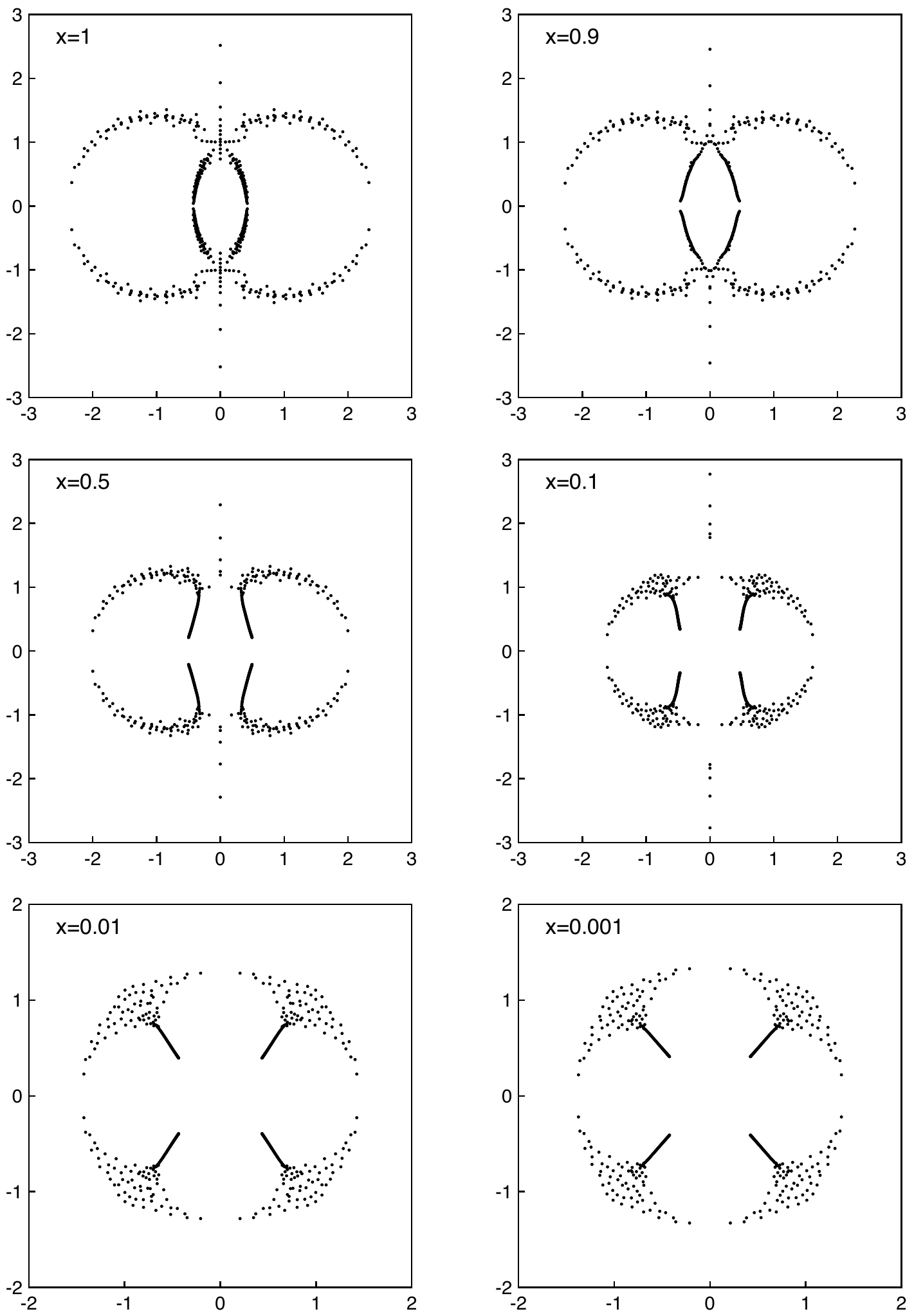,width=9.5cm,angle=0}}
\end{picture}
}
\end{center}
\caption{The zeros in the plane of $y=ux^{1/4}$ for the $16\times 17$
  lattice with toroidal boundary conditions for $x=1.0,~0.9,~0.5,~0.1,~0.01,~0.001.$}

\label{fig8a}
\end{figure}

\begin{figure}[ht]

\begin{center}
\hspace{0cm} \mbox{
\begin{picture}(280,500)
\put(0,0){\epsfig{file=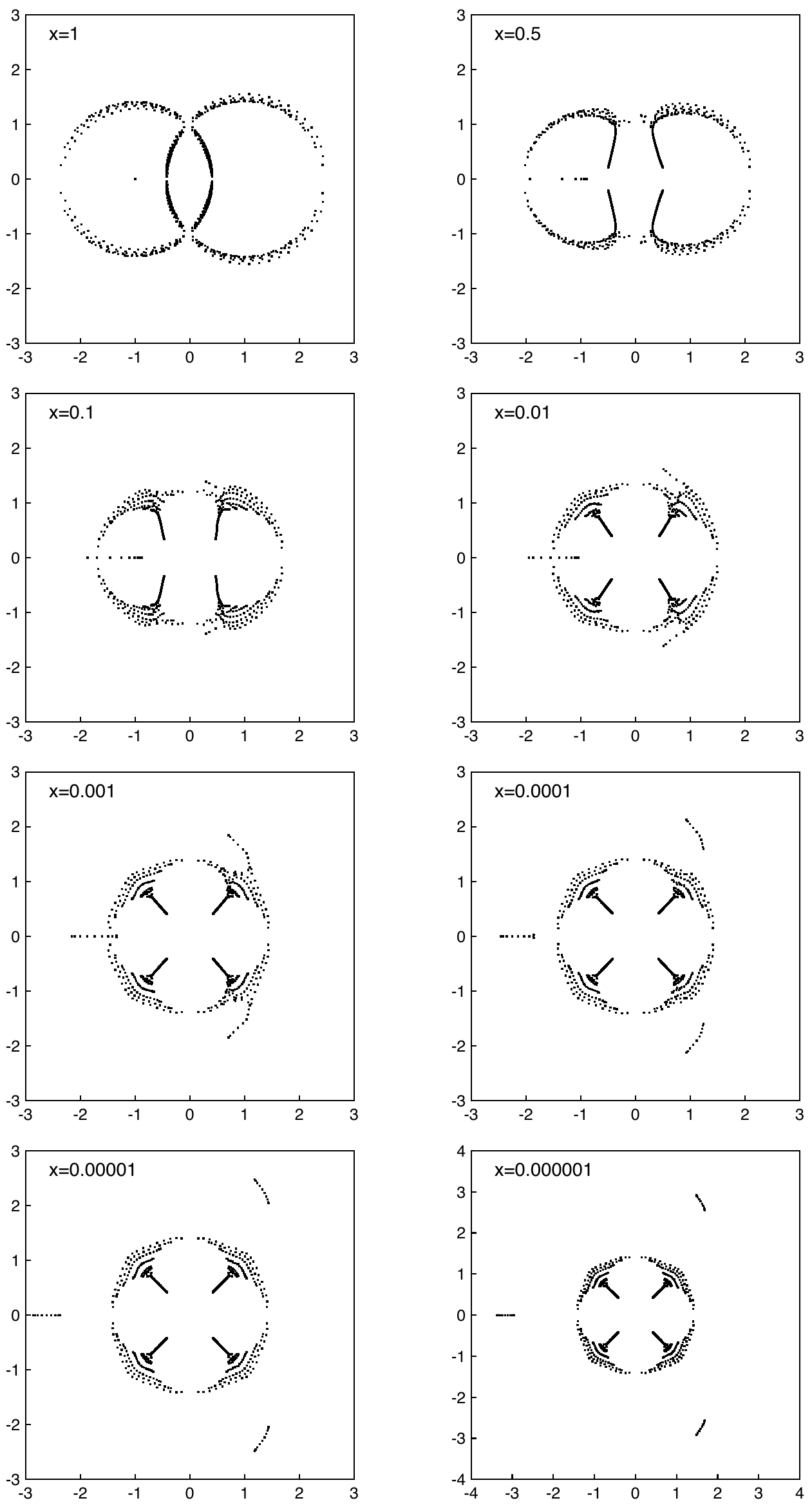,width=9.5cm,angle=0}}
\end{picture}
}
\end{center}
\caption{The zeros in the $y=ux^{1/4}$ plane for the $20\times 20$ 
lattice with cylindrical boundary conditions for $x=1.0,~0.5,~0.1,
~0.01,~0.001,~0.0001,~0.00001,~0.000001.$}

\label{fig8}
\end{figure}

In order to better understand the role on boundary conditions we plot
the zeros as a function of $H$ in the $y=ux^{1/4}$ plane for toroidal
boundary conditions on the $16\times 17$ lattice in Figure~\ref{fig8a} and for
cylindrical boundary conditions of the $20\times 20$ lattice in
Figure~\ref{fig8}.

For cylindrical boundary conditions the exact partition function on
the finite lattice was computed in 1967 \cite{mccoy2}. In contrast
with Brascamp-Kunz boundary conditions the zeros are not symmetric
under $u\rightarrow -u$ and at $u=-1$ the $L\times L$ lattice has an
$L$ fold zero. The total number of zeros is $2L^2-L$.

 As $H$ increases from $H=0$ the $L$ fold zero at $u=-1$ of the $L\times L$
   lattice becomes $L$ zeros on the negative axis which for $L$ even 
are in closely
   spaced pairs. As $H$ is increased the pairs coalesce and become
   complex conjugate pairs.   
   For sufficiently large $H$ they are all complex. However, the
   imaginary part is sufficiently small that in the plots they appear
   to be on the negative axis.

 When $x$ is sufficiently small the  three groups of $L$  zeros each tend to
   infinity at angles $\pi,~\pm \pi/3$. This has previously been seen
   in \cite{kim}.  We
   have no explanation for this phenomenon. The remaining $2L^2-L-3L$
   zeros have a 4-fold symmetry (for L even) at $x\rightarrow 0$.

\section{Transfer matrix eigenvalues}

An alternative method to compute partition functions is to define a
(row to row) transfer matrix on the $L_v\times L_h$ lattice of 
size $2^{L_h}\times 2^{L_h}$. We denote by 
$T_C(L_h)$ the transfer matrix with periodic boundary conditions 
in the $L_h$ direction and by
$T_F(L_h)$ the transfer matrix with free boundary conditions 
in the $L_h$ direction.

In 1949 Kaufman \cite{kauf} computed all eigenvalues of $T_C(L_h)$ and
found that there are two sets 
\begin{equation}
\lambda_{+}=\prod_{n=0}^{L_h-1}e^{\pm \gamma_{2n+1}}\hspace{.2in}
\lambda_{-}=\prod_{n=0}^{L_h-1}e^{\pm \gamma_{2n}}
\label{kauf1}
\end{equation}
 with
\begin{equation}
 e^{\pm\gamma_m}=s+s^{-1}-\cos \phi_m
\pm \left( (s+s^{-1}-\cos \phi_m)^2-1\right)^{1/2}
\label{kauf2}
\end{equation}
where $\phi_m=\pi m/L_h$
and there must be an even number of minus signs.
Each set of eigenvalues contains $2^{L_h-1}$ eigenvalues.

For all $\gamma_{m}$ for $m\neq 0$ the square roots  are defined 
as positive for $0<T<T_c~~(1<s<\infty)$. 

For $|s|=1$ and all $\phi_m$ such that $(s+s^{-1}-\cos \phi_m)^2<1$ 
the modulus of
$e^{\pm \gamma_m}$ is unity and thus many eigenvalues on the circle
$|s|=1$  will have the same modulus.  

For $\gamma_0$  a
factorization occurs under the square root and 
\begin{equation}
e^{\gamma_0}=s+s^{-1}-1+(s-1)(s^{-2}+1)^{1/2}
\end{equation}
So  $\gamma_0$ is positive for $s>1$ and
negative for $s<1$. For $T=T_c$ we have $s=1$ and $\gamma_0=0$. 

There are four constructions of partition functions from these
transfer matrices.
\begin{itemize}
\item $L_v$ periodic, $L_h$ periodic
\begin{equation}
Z^{CC}_{L_v,L_h}={\rm Tr}T_C(L_h)^{L_v}=\sum_k\lambda^{L_v}_{C;k}(L_h),
\end{equation}
\item $L_v$ periodic, $L_h$ free
\begin{equation}
Z^{C,F}_{L_v,L_h}={\rm Tr}T_F(L_h)^{L_v}=\sum_k\lambda^{L_v}_{F;k}(L_h)
\end{equation}
\item $L_v$ free, $L_h$ periodic
\begin{equation}
Z^{FC}_{L_v,L_h}={\bf v}\cdot T_C^{L_v-1}(L_h){\bf v'}
=\sum_k{\bf v\cdot v_k}\lambda_{C;k}^{L_v-1}{\bf v_k\cdot v'}
\end{equation}
\item $L_v$ free, $L_h$ free
\begin{equation}
Z^{FF}_{L_v,L_h}={\bf v}\cdot T_F^{L_v-1}(L_h){\bf v'}
=\sum_k{\bf v\cdot v_k}\lambda_{F;k}^{L_v-1}{\bf v_k\cdot v'}
\end{equation}
\end{itemize}
where $\lambda_{C;k}$ and $\lambda_{F;k}$ are eigenvalues, 
${\bf v}$ and ${\bf v'}$ are suitable boundary vectors and
${\bf v_k}$ are the eigenvectors.

It is obvious by symmetry that
  $Z^{CF}_{L_h,L_v}=Z^{FC}_{L_v,L_h}$ and thus the explicit results of
  1967 for $Z^{FC}_{L_v,L_h}$ must be obtainable from the eigenvalues
  of $T_F(L_h)$ but the eigenvalues of $T_F(L_h)$ have never been
  computed. Clearly something is missing.

\subsection{Equimodular curves}

The Ising model at $H=0$ and $H/k_bT=i\pi/2$  are the only models 
where the finite size partition
function (at arbitrary size) has ever been computed from the transfer matrix eigenvalues.
For all other models when there is one eigenvalue $\lambda_{\rm max}$ 
that is dominant (i.e. of maximum modulus) on the finite lattice the free energy per site in 
the thermodynamic limit is computed as
\begin{equation}
-F/kT=\lim_{L_h\rightarrow \infty}\lim_{L_v\rightarrow
  \infty}\frac{1}{L_vL_h}\ln \lambda^{L_v}_{\rm max}(L_h). 
\end{equation}
However an eigenvalue which is dominant  in one portion of the 
$u=e^{-2E/kT}$ plane will not, in general, be dominant in all parts of
the plane. The places where two or more eigenvalues have the same modulus form
equimodular curves and can separate the complex $u$ plane into
many distinct regions.

When there are only two equimodular eigenvalues $\lambda_1(L_h)$ and
$\lambda_2(L_h)$ on the equimodular curve and there are periodic
boundary conditions in the $L_v$ direction we can approximate the
partition function near the curve as
\begin{equation}
Z_{L_v,L_h} \sim \lambda_1(L_h)^{L_v}+\lambda_2(L_h)^{L_v}  
\end{equation}
and thus for fixed $L_h$ as $L_v\rightarrow \infty$ there will
be a smooth distribution of zeros with a spacing of $1/L_v$ and a
density determined by the phase difference between the two
eigenvalues \cite{mccoy1}.

 For free boundary conditions we have
\begin{equation}
Z_{L_v,L_h} \sim c_1\lambda_1(L_h)^{L_v}+c_2\lambda_2(L_h)^{L_v}  
\end{equation}
where $c_j={\bf(v\cdot v_j)(v_j\cdot v')}$

When there are only two equimodular eigenvalues this relation for
zeros is sufficient for partition functions computed
by first taking $L_v\rightarrow\infty$ and then taking $L_h\rightarrow
\infty$ so that the aspect ratio $L_h/L_v$ vanishes.
For thermodynamics to be valid the free energy must be independent of
aspect ratio as long as $0<L_h/L_v<\infty$. 

\subsection{Equimodular curves for $T_C(L_h)$ at $H=0$}

\begin{figure}[ht]

\begin{center}
\hspace{0cm} \mbox{
\begin{picture}(360,180)
\put(0,0){\epsfig{file=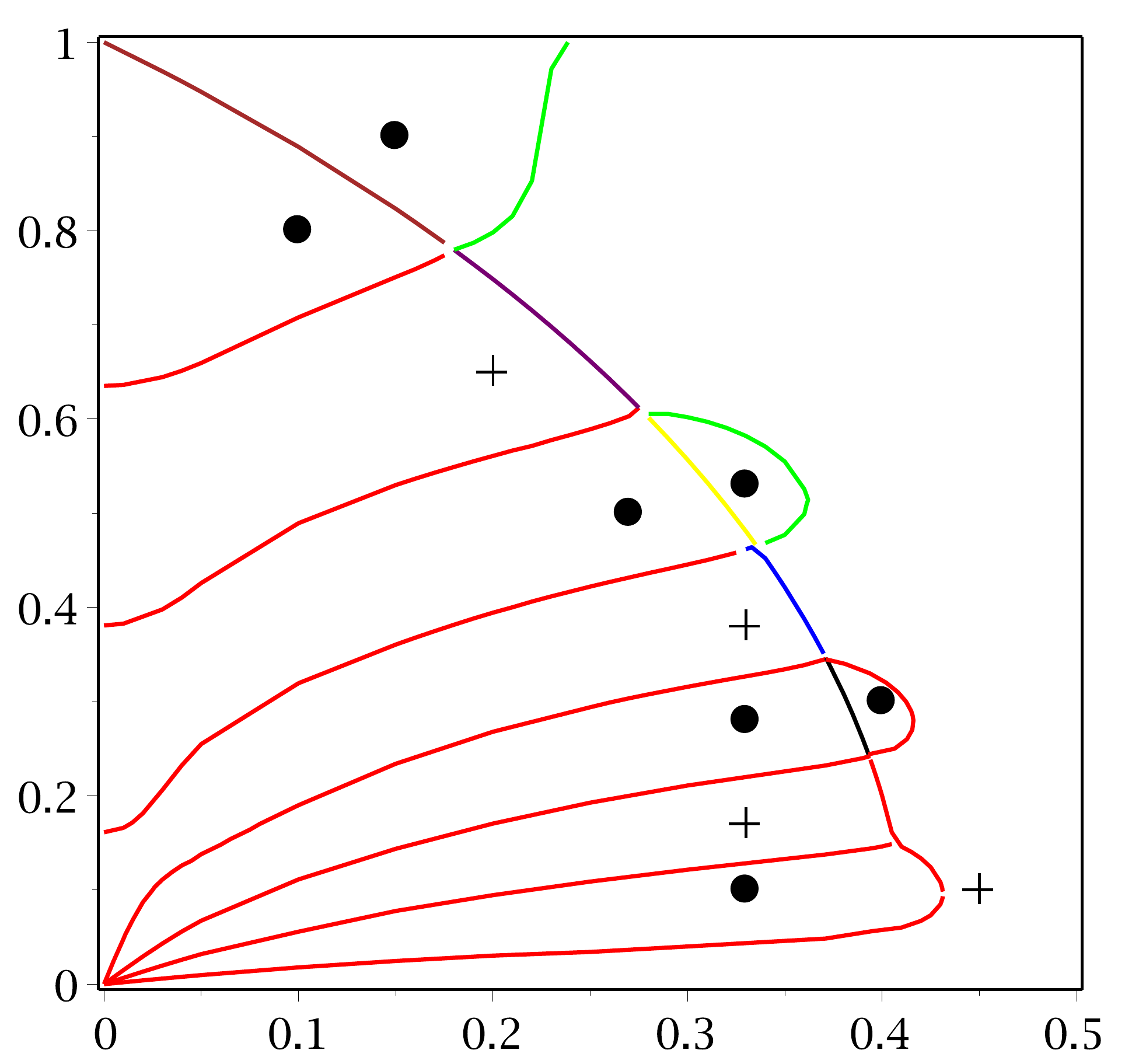,width=6cm,angle=0}}
\put(180,0){\epsfig{file=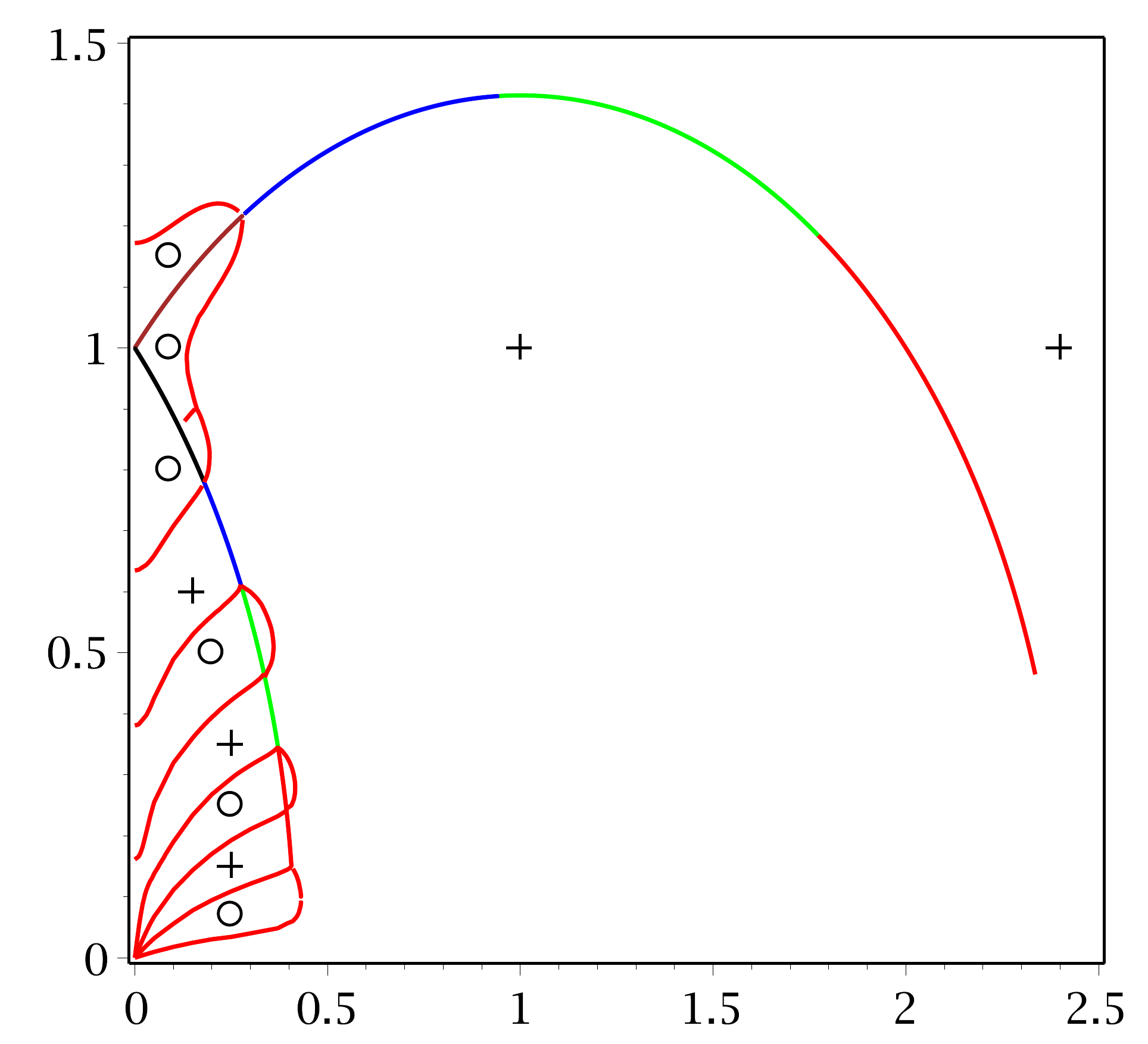,width=6.2cm,angle=0}}
\end{picture}
}
\end{center}
\caption{The equimodular curves in the $u$ plane for $T_C(L_h)$ 
for $L_h=8$. On the
  left all eigenvalues are considered and on the right the restriction
  to the momentum sector $P=0$ is made. The sectors where $\lambda_{+}$ is
  dominant is marked by $+$ and the sector where $\lambda_{-}$ is
  dominant is marked by a circle. The multiplicity of the crossings on
  the curves are indicated by colors. On left panel:red=2,  
green=3, black=4, blue=8, yellow=16, purple=32, brown=64
On right panel: red=2,  green=4,  blue=8,
  brown =3, black=9.}
 \label{fig9}
\end{figure}

\begin{figure}[ht]

\begin{center}
\hspace{0cm} \mbox{
\begin{picture}(360,180)
\put(0,0){\epsfig{file=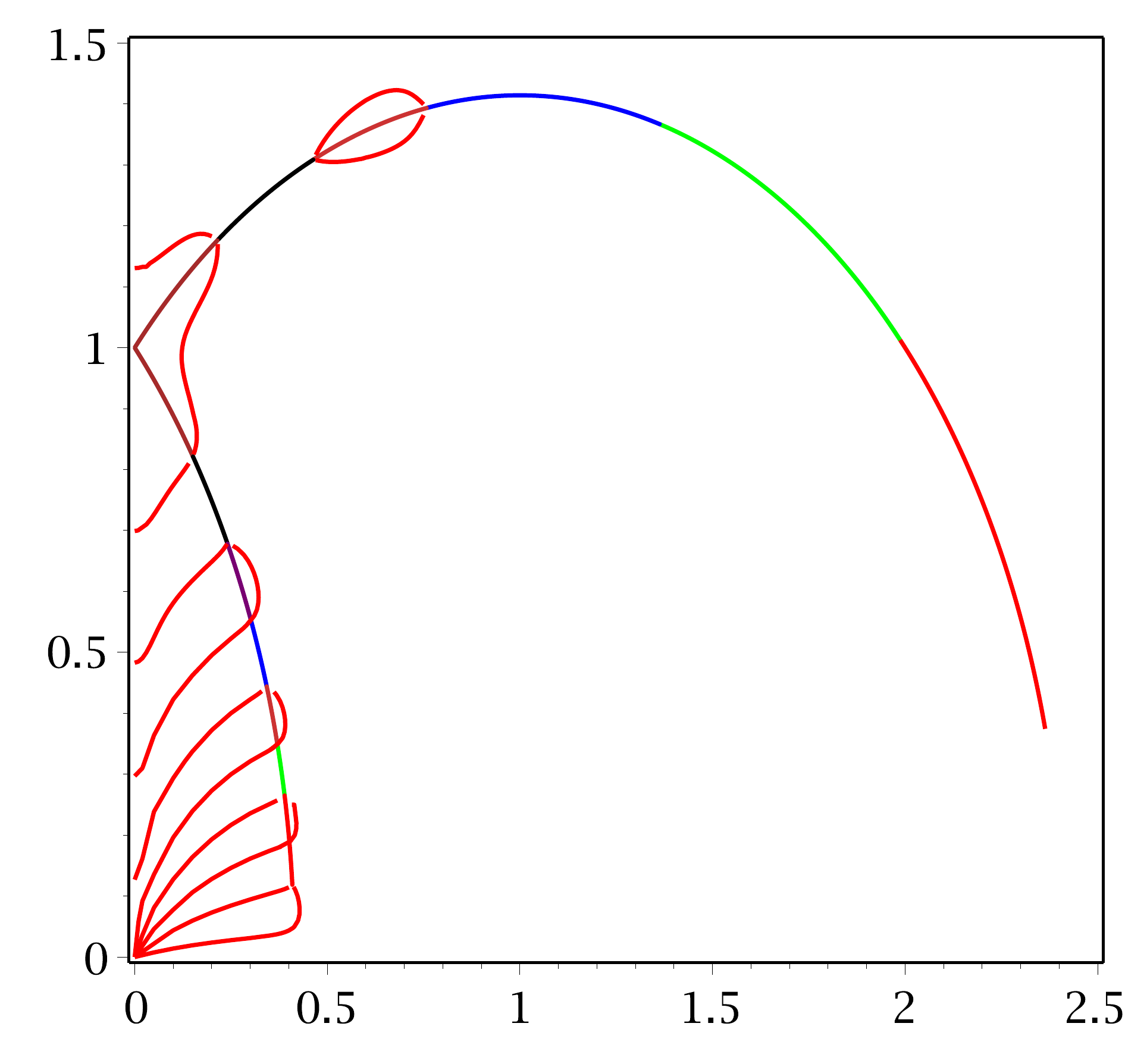,width=6cm,angle=0}}
\put(180,0){\epsfig{file=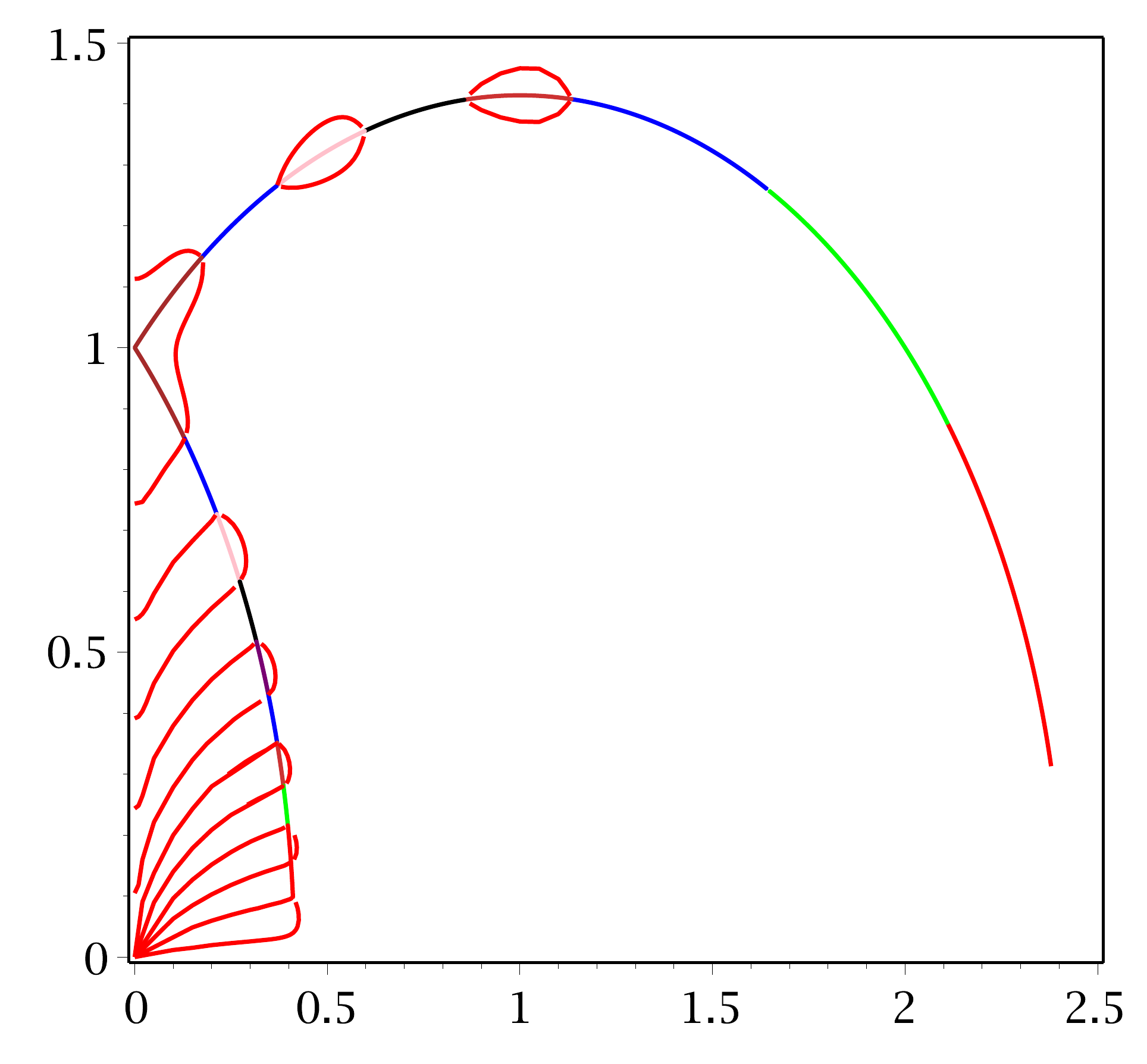,width=6cm,angle=0}}
\end{picture}
}
\end{center}
\caption{The equimodular curves in the $u$ plane for $T_C(L_h)$ 
at $P=0$ for $L_h=10$ on
  the left and 12 on the right. Red indicates a multiplicity of 2,
  green of 4 and blue of 8. 
For $L=10$ the sequence of multiplicities on the upper 
(antiferromagnetic) sequence (increasing towards $u=i$) is  2,4,8,4,18,24 
and the lower (ferromagnetic) sequence 2,2,4,4,8,8,18,28.
For $L=12$ the upper sequence 2,4,8,2,18,18,52,84 and the
lower sequence is 2,2,4,4,8,8,18,26,52,88}
\label{fig10}
\end{figure}

For the Ising model at $H=0$ the equimodular curves of the transfer
matrix $T_C(L_h)$ can be numerically  computed from the eigenvalues
(\ref{kauf1}),(\ref{kauf2})
of Kaufman \cite{kauf} where we note that the corresponding momentum
is
\begin{equation}
P=\sum_{m}\phi_m~~({\rm mod}~2\pi).
\end{equation}
We plot these curves in the complex $u$ plane 
in Figures~\ref{fig9} and \ref{fig10} for $L_h=8,10,12$.

\clearpage

\clearpage

These curves have the following striking properties:
\begin{enumerate}

\item  All eigenvalues are equimodular at $u=\pm i$. 

\item  The equimodular curves in the $u$ plane of the eigenvalues 
$\lambda_{+}$ and the  eigenvalues $\lambda_{-}$ are segments of the
two circles $u=\pm 1+2^{1/2}e^{i\theta}$
which is the curve on which there are Brascamp-Kunz zeros.

\item  On most of the segments of this curve there are more than two
   equimodular eigenvalues.

\item  The equimodular curves formed by one eigenvalue $\lambda_{+}$ and
   one $\lambda_{-}$ do not lie on the curve of Brascamp-Kunz zeros.
\end{enumerate}
The multiple degeneracies on the equimodular curves destroy the
mechanism for a smooth density of zeros of the $L_v=L_h=L$ lattice 
with a $1/L^2$ spacing. The mechanism which changes the scale of 
smooth zeros from $1/L^2$ to $1/L$ seen in section 3.3 is not understood.

 \subsection{$u$ plane eigenvalues for $x=0.99$}

\begin{figure}[ht]

\begin{center}
\hspace{0cm} \mbox{
\begin{picture}(400,120)
\put(0,125){\epsfig{file=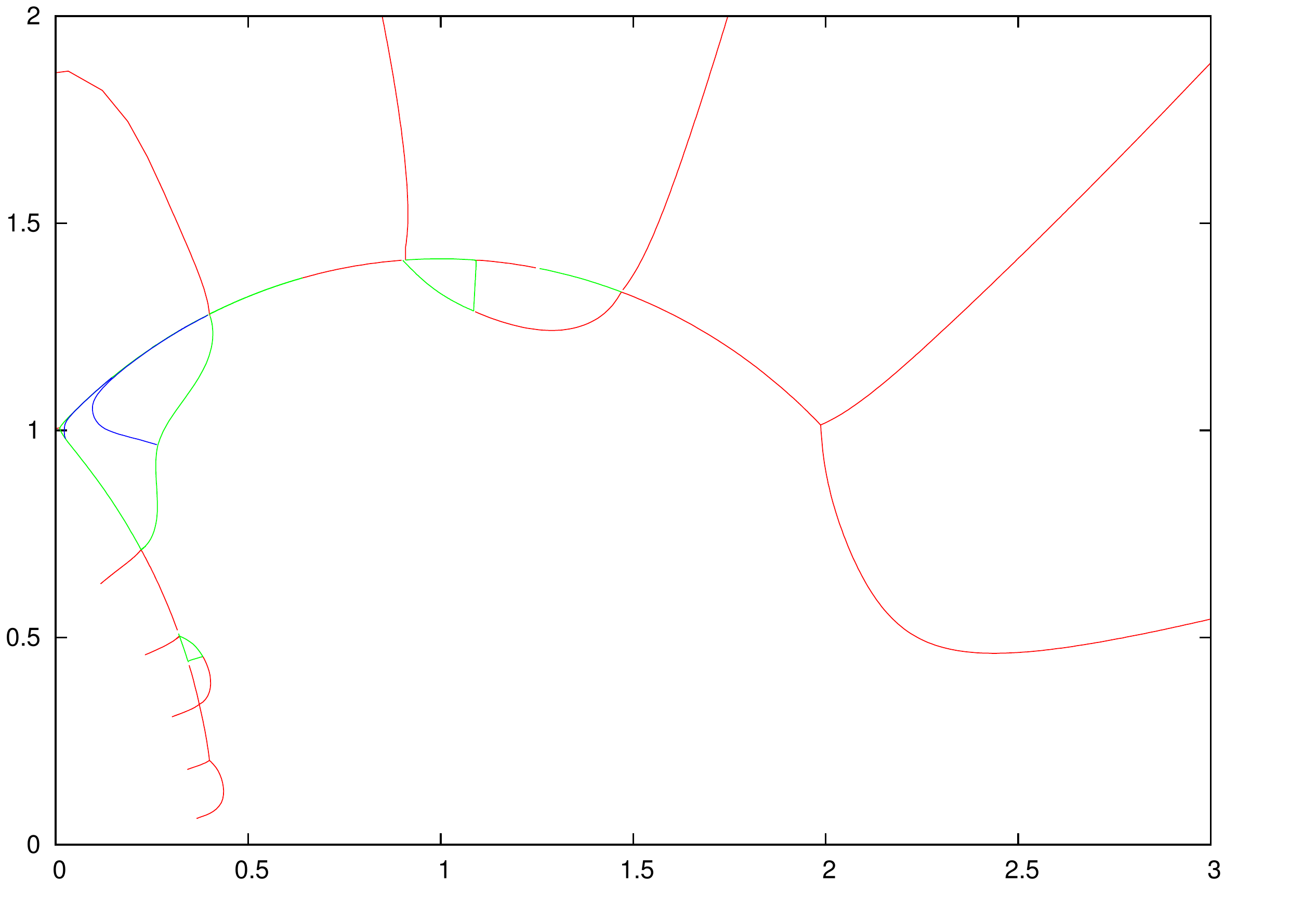,width=4.5cm,angle=-90}}
\put(200,125){\epsfig{file=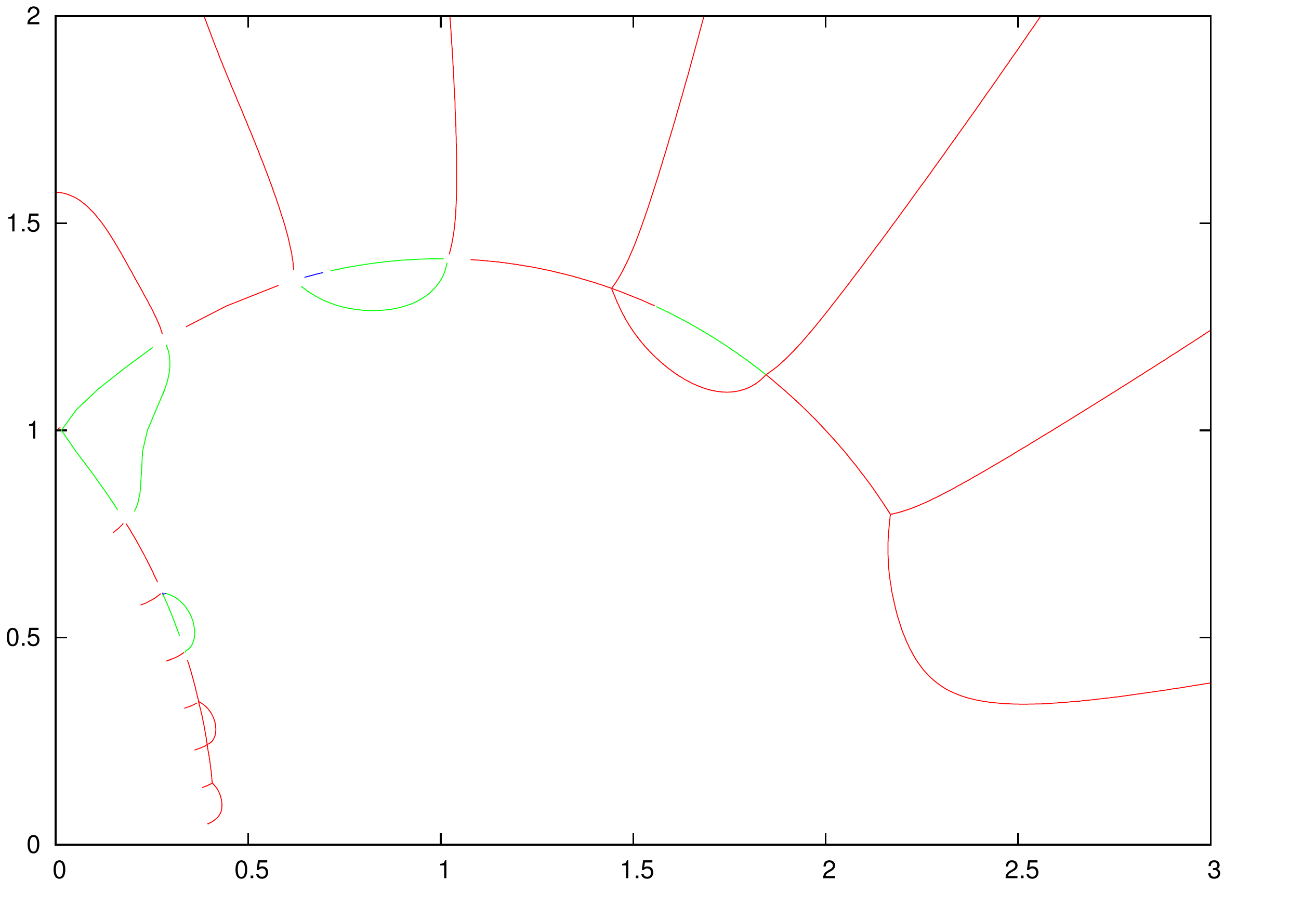,width=4.5cm,angle=-90}}
\end{picture}
}
\end{center}
\caption{Equimodular curves in the $u$ plane for $x=0.99$ of
  $T_C(L_h)$ for $L_h=6$ on the left and $L_h=8$ on the right. 
 Red is for singlet-singlet crossings,
 green is for singlet-doublet and 
 blue is for doublet-doublet}
\label{fig11}
\end{figure}

When $H$ is increased from $H=0$ the transfer matrix eigenvalues have been 
computed numerically.  In Figure~\ref{fig11} we plot the equimodular
curves for  all eigenvalues for $x=0.99$. 
(We note that the curves extending from the upper
branch to infinity are also present for $H=0$ but are not seen in
Figure~\ref{fig9} because in that figure the imaginary part of $u$ is
restricted to $0\leq \mathrm{Im}(u)\leq 1$.)

By comparing Figure~\ref{fig11} with Figures~\ref{fig9} and \ref{fig10}
we see that several dramatic phenomena occur for $H>0$.

\begin{enumerate}

\item For $H>0$ the rays to the imaginary axis very rapidly retreat into
   the curve of the Brascamp-Kunz zeros. This is caused by the 
lifting of the near degeneracy of
   eigenvalues in the $\lambda_+$ and $\lambda_{-}$ subspaces of
   $H=0$. The larger $L_h$ the more rapid the retreat.

\item The rays to infinity separate regions of $P=0$ and $P=\pi$ and  are
virtually unchanged for $H>0$.

\item The multiple degeneracies disappear. For momenta $P=0,\pi$ 
the eigenvalues are singlets for $P\neq 0,~\pi$ the  momenta $\pm P$ 
are doubly degenerate. In Figure~\ref{fig11} all singlet-doublet 
and doublet-doublet curves enclose regions
where the dominant eigenvalue has $P\neq 0,\pi$ but for $x=0.99$  some of the
regions are too small  to be observed as areas.

\end{enumerate}

\begin{figure}[ht]

\begin{center}
\hspace{0cm} \mbox{
\begin{picture}(200,180)
\put(0,180){\epsfig{file=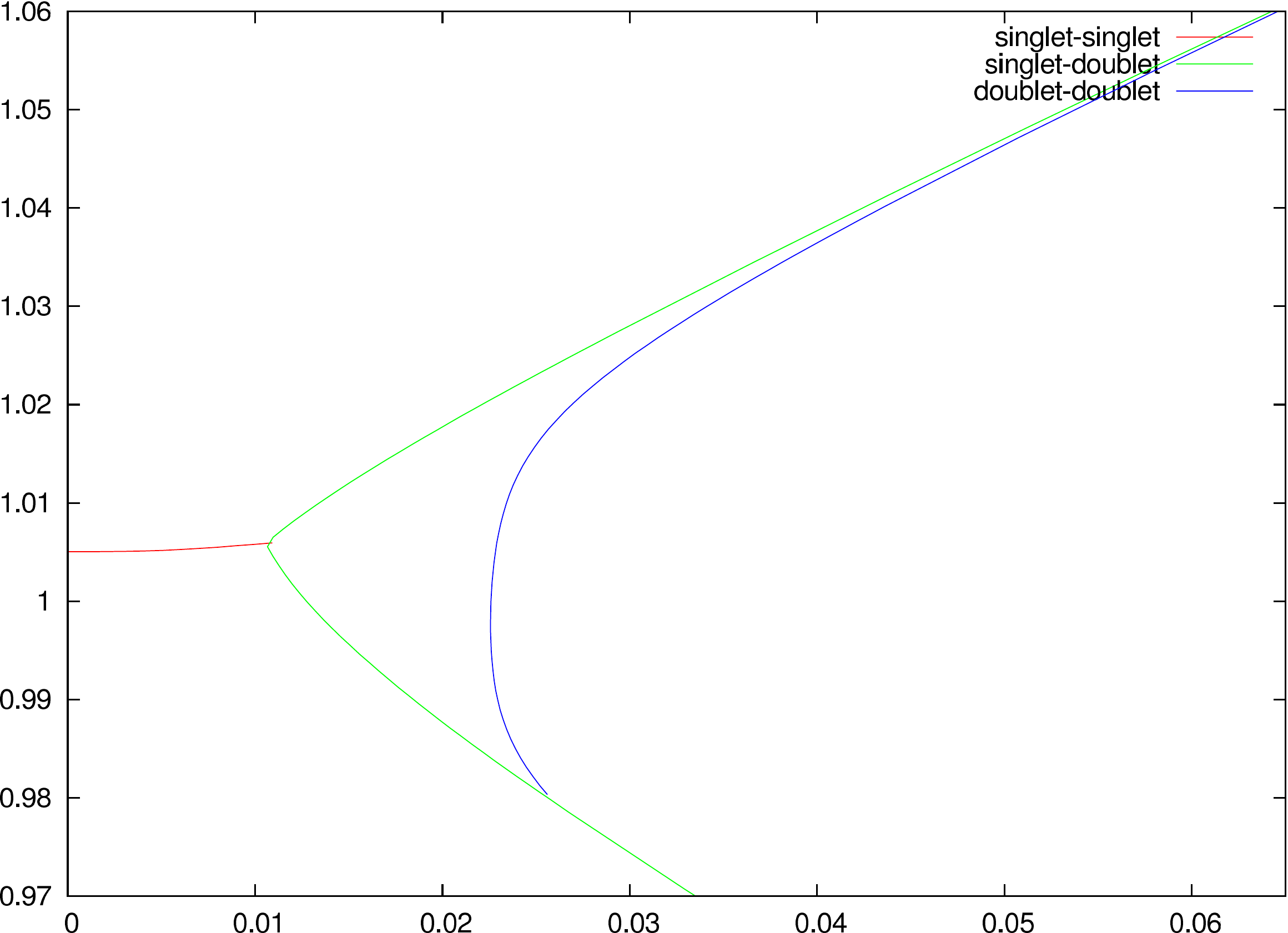,width=6.5cm,angle=-90}}
\end{picture}
}
\end{center}
\caption{Equimodular curves in the $u$ plane for $x=0.99$ expanded near 
$u=i$ for $T_c(L_h)$ with $L_h=8$. Red is for singlet-singlet crossings,
green is for singlet-doublet and blue is for doublet-doublet}
\label{fig12}
\end{figure}

In Figure~\ref{fig12} we plot for $L_h=8$ the region near $u=i$ in more detail.
Thus far eigenvalues for $L_h\geq 10$ have not been computed for the
case $H\neq 0$.

\section{An interpretation} 

It is very clear, both from the behavior of the partition function
zeros and the degeneracy of the equimodular curves, that there is a
drastic qualitative difference between $H=0$ and $H\neq 0$. We
conjecture here an interpretation of the
singularities (\ref{nickel}) found by 
Nickel \cite{nickel1,nickel2} based on
this behavior. The argument is substantially different for the inner
(ferromagnetic) and outer (antiferromagnetic) loops in the $u$ plane.
Naturally conjectures concerning analyticity based solely on finite
size computations can only be suggestive.

\subsection{Scenario on the ferromagnetic loop}

We conjecture that on the ferromagnetic loop for $H>0$ the zeros
approach a curve as $L_hL_v=N\rightarrow \infty$ and that for
sufficiently large $N$ and fixed $H\neq 0$ the limit
\begin{equation}
\lim_{N\rightarrow \infty}N(u_{j+1}-u_j)<\infty
\end{equation}
exists. However, this cannot be uniform in $H$ and thus the limits
$H\rightarrow 0$ and $N\rightarrow \infty$ will not commute.
For both $H=0$ and $H\neq 0$ the free energy is analytic at the locus
of zeros. However, for $H\neq 0$ the analytic continuation beyond the
zero locus encounters many singularities which accumulate in the limit
$H\rightarrow 0$ to the singularities of Nickel (\ref{nickel}). The
location (and nature) of these singularities is different if the
continuation is from the interior (low temperature) or exterior (high
temperature)  of the loop. The amplitude of the singularities vanishes
as $H^2$ at $H\rightarrow 0$ and hence the analyticity of the free
energy at $H=0$ is maintained. 

In this scenario the singularities in the susceptibility at $|s|=1$ occur
because taking two derivatives with respect to $H$ kills the $H^2$
in the amplitude of the singularities but does not move the locations.

It can be argued that the non-integrability of the Ising model at
$H\neq 0$ is caused by these singularities in the analytic
continuation beyond the locus of zeros. Nevertheless, there are no
singularities on the locus of zeros except at the endpoints. The singularity
at the endpoint is expected \cite{mccoy1} to have the 
same behavior as the endpoint
behavior of hard squares, hard hexagons and the Lee-Yang edge. 

We may now make contact with the scenario of Fonseca and
Zamolodchikov \cite{fz} who assume that in the field theory limit the
free energy may be continued far beyond the locus of zeros.
The field theory limit is defined by $T\rightarrow T_c$ and
$H\rightarrow 0$ such that
\begin{equation}
\tau=(T-T_c)H^{-8/15}
\end{equation}
is fixed of order one. In terms of this scaled variable Fonseca and
Zamolodchikov posit that there is analyticity across the locus of
zeros and that there is an extensive region of analyticity in the
analytically continued free energy which sees none of the 
singularities which, in this interpretation, produce the singularities
of Nickel. The analyticity of \cite{fz} will be consistent with our scenario
 if the singularities which approach the  point 
$u={\sqrt 2}-1$ as $H\rightarrow 0$ is slower than the scaling $H^{8/15}$. 
If this is indeed the case then there is no contradiction between the
field theory computations of \cite{fz} and the singularities 
of \cite{nickel1,nickel2}.

\subsection{Scenario on the antiferromagnetic loop}

The behavior on the antiferromagnetic loop is quite different from
the behavior on the ferromagnetic loop because
now the zeros spread out into areas for $H\neq 0$. Moreover the pinching of the
zeros at the antiferromagnetic singularity at $u={\sqrt 2}+1$ remains
a pinch for all values of $x$ and furthermore the singularity in the
free energy in the hard square limit is numerically estimated from
high density series expansions
\cite{baxter2,kb} to be the same as the logarithmic singularity  at 
$T_c$ of the antiferromagnetic Ising model at $H=0$.  

The zeros in
Figures~\ref{fig4} and \ref{fig5} do appear to be smoothly spaced in a two
dimensional region so from this point of view the distribution of zeros
which for $H=0$ was studied in section 3.3 has moved smoothly from
the circle to an area in the plane. There is, unfortunately,  not sufficient
data to conjecture the behavior where the zeros in the $N\rightarrow
\infty$ limit pinch the positive $u$ axis. Even in the hard square
limit it cannot be concluded from Figure~\ref{fig5} if the zeros pinch
as a curve, as a cusp with an opening angle of zero or as a wedge with
a nonzero opening angle. The  field theory argument of
\cite{fz}  does not extend to the hard square limit and it is not 
obvious how to consider analytic continuation into an area of zeros.

The second feature which needs an explanation is the 
approach of the zeros to the hard square limit 
in both Figure~\ref{fig4} for Brascamp-Kunz 
boundary condition in the $y=u^2x^{1/2}$ plane and in
Figures~\ref{fig8} and \ref{fig8a} 
for cylindrical and toroidal boundary conditions in the $y=ux^{1/4}$
plane. Namely
the emergence of the 2 fold symmetry for Brascamp-Kunz and the 4 fold
symmetry for cylindrical and toroidal boundary conditions. For all boundary
conditions new points of singularity are created in the complex $y$
plane as $H$ is increased, which in the hard square limit become identical
with the singularity on the positive $y$ axis. The mechanism for
the creation of these new points of singularity is completely unknown.

\subsection{The bifurcation points}

However, perhaps the most striking feature of the zeros is the existence 
of the special points where the
one dimensional locus bifurcates into the two dimensional area. 
It is the existence of these points which allows us to use the terms
ferromagnetic and antiferromagnetic branch. 
At $H=0$ these points are  at $u=\pm i$ where all eigenvalues are
equimodular and the free energy is singular \cite{shrock1}. In the hard square 
limit this point is at $z=-1$ where
all eigenvalues are also equimodular \cite{fen}. It is natural to
conjecture that for all values of $H$ the free energy fails to be 
analytic at these points. 

\section{Conclusion}

In this paper we have presented the results of extensive numerical
computations of the zeros of the partition function of the Ising model
in a magnetic field $H$ and a companion study of the dominant
eigenvalues of the transfer matrix as  
$H$ goes from $H=0$ to the hard square limit $H\rightarrow \infty$.
This reveals that in the ferromagnetic region the distribution of
zeros changes radically when $H$ is infinitesimally increased from $H=0$
and this  feature is used to give an interpretation of the natural
boundary in the magnetic susceptibility conjectured by Nickel 
\cite{nickel1,nickel2} which is consistent with the analyticity of
the scaling limit assumed by Fonseca and Zamolodchikov
\cite{fz}. However, an analytic argument for this scenario remains to
be found and  further data is needed in order to reliably
understand the approach to the hard square limit.

\section*{Acknowledgments}

We are pleased to thank for their hospitality the organizers of 
the conference ``Exactly solved models and beyond'' held at 
Palm Cove, Australia July 19-25, 2015 in honor of the 75th birthday 
of Prof. Rodney Baxter where much of this
material was first presented. One of us (JLJ) was supported by the
Agence Nationale de la Recherche (grant ANR-10-BLAN-0414), the
Institut Univeritaire de France, and the European Research Council
(advanced grant NuQFT). Two of us (MA and IJ) were supported by
funding under the Australian Research Council's Discovery Projects
scheme by the grant DP140101110. The work of IJ was also supported by
an award under the Merit Allocation Scheme of the NCI National Facility.

\end{document}